\documentclass[preprint,nofootinbib,aps,superscriptaddress,eqsecnum]{revtex4-1}
 \pdfoutput=1
\usepackage{amsmath,amssymb,graphicx,subfigure}

\usepackage{float}
\usepackage{graphicx}
\usepackage{amsmath}
\usepackage{amsfonts}
\usepackage{amssymb}
\usepackage{hyperref}
\usepackage{subfigure}
\usepackage{slashed}
 \usepackage{setspace} 
\usepackage{caption}
\usepackage{color}
\allowdisplaybreaks
\def\bea{\begin{eqnarray}}
\def\eea{\end{eqnarray}}
\def\be{\begin{equation}}
\def\ee{\end{equation}}
\def\nn{\nonumber}

\begin{document}

\title{A new feasible dark matter region in the singlet scalar scotogenic model}

\author{Pritam Das}
\email{pritam@tezu.ernet.in}
\affiliation{ Department of Physics, Tezpur University, Assam-784028, India}
\author{Mrinal Kumar Das}
\email{mkdas@tezu.ernet.in}
\affiliation{ Department of Physics, Tezpur University, Assam-784028, India}
\author{Najimuddin Khan}
\email{psnk2235@iacs.res.in}
\affiliation{Centre for High Energy Physics, Indian Institute of Science,
	C. V. Raman Avenue, Bangalore 560012, India}
\affiliation{School of Physical Sciences, Indian Association for the Cultivation of Science
	2A $\&$ 2B, Raja S.C. Mullick Road, Kolkata 700032, India}
\begin{abstract}
We study a simplest viable dark matter model with a real singlet scalar, vector-like singlet and a doublet lepton. We find a considerable enhancement in the allowed region of the scalar dark matter parameter spaces under the influence of the new Yukawa coupling.
 The Yukawa coupling associate with the fermion sector heavily dominant the dark matter  parameter spaces satisfying the current relic density of the Universe. Dilepton$+\slashed{E}_T$ signature arising from the new fermionic sector can observe at Large Hadron Collider (LHC). We perform such analysis in the context of 14 TeV LHC experiments with a future integrated luminosity of 3000 ${\rm fb^{-1}}$.
We found that a large region of the parameter spaces can be probed by the LHC experiments.
The projected exclusion/discovery reach of direct heavy charged fermion searches in this channels is analyzed by performing a detailed cut based collider analysis.
The projected exclusion contour reaches up to $1050-1380~{\rm GeV}$ for 3000 ${\rm fb^{-1}}$ for a light dark matter $\mathcal{O}(10)$ GeV from searches in the $ pp \rightarrow E_1^\pm E_1^\mp, E_1^\pm\rightarrow l^\pm S \rightarrow ll + \slashed{E}_T$  channel.

\end{abstract}

	\maketitle

\section{Introduction}
Pieces of evidence from various astrophysical observations like gravitational lensing effects in the bullet cluster, anomalies in the galactic rotation curves have confirmed the existence of dark matter (DM) in the Universe. Since the SM does not have enough particle to play the role of  DM, we must go beyond the SM in search of new physics. 
The recent LHC Higgs signal strength data~\cite{Sirunyan:2017khh, Sirunyan:2018koj} also suggests that one can have rooms for the new physics beyond the SM.
In order to address DM within BSM, various possibilities have been proposed in~\cite{Tanabashi:2018oca,ArkaniHamed:2008qn} and references therein. Extension of the SM with new fields is widespread in the literature, under which the lightest and stable particle due to the imposed discrete $Z_n$ and/or $Z_n$-type ($n\ge 2, integer$) plays the role of dark matter \cite{Dasgupta:2014hha,Babu:2009fd}.
Rich literature on minimal models of DM considering scalar and fermion multiplets are available today~\cite{Burgess:2000yq,Deshpande:1977rw,Ma:2008cu,Araki:2011hm,Ma:2006km}. In particular, the addition of singlet scalar and fermion singlet, as well as doublet in a minimal model, have rich demand in DM study. The mixing of fermion doublet and singlets reduces the coupling to weak gauge bosons. This transform DM from a Dirac into a Majorana particle, yielding the correct relic density with allowed direct detection cross-section~\cite{Cohen:2011ec}.

It is known that the direct detection experiments eliminate a significant portion of the parameter space in the dark matter mass versus dark matter-nucleon cross-section plane. However, these experimental constraints are marred by the uncertainties stemming from the assumption that the Earth is flying through a uniform dark matter cloud of significant density.
	The clumpy nature of dark matter leaves open the possibility that the
	density of dark matter in the cosmologically tiny region surrounding
	the Earth, which has not been directly measured so far, is
	very small.
	This makes the option that dark matter may be produced
	directly at a high-energy collider like the LHC/ILC even more
	attractive.
	Weakly interacting massive particles (WIMPs), i.e., the dark matter in this model can indeed be produced by the proton-proton collisions at
	the LHC which escape the detector, leading to the celebrated
	missing energy signal.
	As backgrounds are somewhat better
	understood in a manmade laboratory, it is not unreasonable
	to argue that a collider might be the best bet in revealing the
	true nature of DM particles. It is also  possible to explain the observations in various indirect dark matter detection experiments for some regions (low-mass) of the parameter space. In this paper, however, we do not discuss such details, as such estimations involve proper understanding of the astrophysical backgrounds and an assumption of the dark matter halo profile which contain some arbitrariness.
Concrete experimental signature of existence of dark matter is hitherto unknown, however, recent Xenon-1T experiment~\cite{Aprile:2018dbl} puts stringent bounds on the dark matter portal interaction strength(s). In the basic hypothesis, there exist non-negligible but little interaction between DM with the SM particles which assures that DM is in equilibrium with a thermal bath.
Eventually, it `freeze-out' from the hot plasma of the SM particles, and we can calculate the current relic density of the DM candidate. DM detection experiments indicate that either dark matter may interact with the nucleus very feebly (detection cross-section could reach beyond the line of neutrino floor~\cite{Vergados:2008jp,Boehm:2018sux}) or the interaction is ultimately zero. 
Hence, the dark matter annihilation into the SM particles via $s$-channels may absent. On the other hand, if Nature has only one-component dark matter, then the $H$- and $Z$-bosons portal light dark matter models may not be the right one to give the exact relic density. It is already clear from the literature \cite{Camargo:2019ukv, Restrepo:2015ura, Ahriche:2017iar, Fiaschi:2018rky} that in the presence of other particles one can get the exact relic density via the co-annihilation channels. There may have interactions in such a way that the dark matter can annihilate into the SM particles via $t$- or $u$-channels. This might help to modify the effective annihilation cross-section to give the exact relic density. These types of scenarios can be achieved in the proposed minimal model, which gives the correct dark matter density satisfying the other theoretical and experimental constraints. 

In the model-building prospect, models that can address more SM shortfalls are much appealing and well-motivated also.
A working model is said to be completed when it can simultaneously explain light neutrino observable and dark matter \cite{Ahriche:2017iar,Babu:2009fd,Baumholzer:2018sfb,Baumholzer:2019twf,Bhattacharya:2017sml,Das:2019ntw,Kashiwase:2015pra}.
The framework that is popular in accommodating both dark matter and neutrino mass at loop level is known as the $scotogenic$ model. It was first proposed by E. Ma~\cite{Ma:2006km}, where the dimension-5 operator is realized at the one-loop level. The notable feature of this framework is the way it connects neutrino and DM. Due to the additional $Z_2$ discrete symmetry, new fields that contribute to the loop to produce sizable neutrino mass, acquire opposite parity to the SM fields. Hence, the new field becomes stable and can be addressed as a viable dark matter candidate. Due to its impressive features in addressing neutrino mass and dark matter, the scotogenic model has gained popularity over time~\cite{GonzalezFelipe:2003fi,Fraser:2014yha, Merle:2015ica, Law:2013saa,Mahanta:2019gfe, Klein:2019iws}. In order to get relic density of the singlet scalar dark matter and allowed from the direct detection, one has to take the dark matter particle to be heavy or else take the Higgs portal couplings to be small near the Higgs resonance region.
	This spoils the testability of the theory at the collider. Hence, one need to extend the model model with a vector-like fermions (to cancel effect of the anomaly), which can be probed at the collider
	experiments. In this model we found a large region of the dark matter parameter spaces in the presence of the vector-like fermions allowed by the all theoretical and experimental constants. Here, we have not only introduced a viable dark matter candidate but also comment on the possible explanation of tininess of neutrino mass generation under a single framework with minimum particle content. Initially, we will be starting with just a single generation of fermion doublet that interacts with the lepton doublet and check whether it is sufficient to address both the neutrino parameters and dark matter or not.

Keeping these in view, we consider a minimal model of DM comprise of a vector-like singlet and doublet lepton along with a singlet scalar. We introduce an additional $Z_2$ symmetry, under which all new fields are assigned odd, which restricts its interaction with SM particles. 
The viable DM candidate in the extended singlet scalar model is the lightest $Z_2$-odd singlet scalar $S$. 
In the presence of the Yukawa couplings, a considerable improvement to the region of the dark matter parameter space is noticed in this present work. Depending upon the size of the Yukawa couplings, one can get a dominant DM annihilation through $t$- and $u$-channels. The interference between the $s$-channel and cross-channel (2-singlet,2-Higgs scalar vertex), and $t,u$-channels played a crucial role in achieving the correct DM density. The co-annihilation channels also played an essential role in getting a viable region of allowed dark matter parameter space.

The lepton flavour violating processes ($\mu \rightarrow e \gamma$), electron and muon anomalous magnetic moment are also a striking indication of BSM. As there is a discrepancy between the measured value and the SM predictions~\cite{Bennett:2006fi, Parker:2018vye}:
$\delta a_{\mu}=a_{\mu}^\text{exp}-a_{\mu}^\text{SM}=(2.74\pm 0.73)\times 10^{-9}$
and 
$\delta a_{e}=a_{e}^\text{exp}-a_{e}^\text{SM}=-(8.8\pm 3.6)\times 10^{-13}$.
Among the popular works on the discrepancy of the muon magnetic moment, some of them are due to the addition of extra Higgs boson\cite{Abe:2017jqo,Chun:2016hzs}, introducing a light $Z^{\prime}$ gauge boson associated with an extra $U(1)_{L_{\mu}-L_{\tau}}$ symmetry \cite{Baek:2001kca}, or a light hidden photon \cite{Endo:2012hp}, imposing discrete symmetries \cite{Abe:2019bkf}. In those models, the muon magnetic moment is enhanced with a smaller coupling strength via loop mediator process. In this proposed minimal model, we will also try to comment on the discrepancy of the muon anomalous magnetic moment mediated via a vector-like fermion.

In the present paper, we have identified the parameter space relevant to dark matter, lepton flavour violation and neutrino masses. In future, if this type of model turns out to be the dark matter model realized in Nature, our study could help in estimating a better parameter space.
Moreover, the interaction of vector-like fermions with SM fields makes them more comfortable to probe in collider searches \cite{Bhattacharya:2017sml,Bhattacharya:2018cgx,Cynolter:2008ea}. We look for collider signature for the lightest charged fermion in the context of 14 TeV LHC experiments with a future luminosity of 3000 ${\rm fb^{-1}}$ for $pp\rightarrow E_1^{\pm}E_1^{\mp}$ event processes which yield dilepton plus large transverse missing energy $\slashed{E}_T$ (arising from the dark matter) in the final state. 
We get significant results from the collider searches for the discovery of dark matter in future 14 TeV run with the said luminosity.
The projected exclusion/discovery reach of direct heavy charged fermion searches in this channels is analyzed by performing a detailed cut based collider analysis.
The projected exclusion contour reaches up to $1050-1380~{\rm GeV}$ for a light dark matter $\mathcal{O}(10)$ GeV from searches in the $ pp \rightarrow E_1^\pm E_1^\mp, E_1^\pm\rightarrow l^\pm S \rightarrow ll + \slashed{E}_T$  channel. 
To the best of our knowledge, detailed analysis of this model has not yet been done in the literature, which motivates us to carry out the analysis. 

The rest of the work is organized as follows. We have given the complete model description in section~\ref{sec2}. Constraints from various sources on this model are discussed in section~\ref{s3}. Numerical analysis for dark matter, neutrino and collider searches are discussed under section~\ref{s5}. Finally, we conclude our work in section~\ref{conc}. 
\section{Model framework}\label{sec2}
\begin{table}[h!]
	\begin{tabular}{|c|ccc|}
		\hline
		Fields$\rightarrow$ &$S$&$F_D$&$E_S$\\
		Charges$\downarrow$	&&&\\
		\hline
		$SU(2)$&1&2&1\\
		$U(1)_Y$&0&-1&-2\\
		$Z_2$&-1&-1&-1\\
		\hline	
	\end{tabular}
	\caption{Particle content and their charge assignments under $SU(2)$, $U(1)_Y$ and $Z_2$ groups.}\label{modelt}
\end{table}
The model addressed here, contains (i) a real singlet scalar ($S$), (ii) a vector-like charged fermion singlet $E_S^-$ and (iii) a vector-like fermion (VLF) doublet,
 $F_D=(X_1^0~~E_D^-)^T$ \cite{Bhattacharya:2017sml, Bhattacharya:2018cgx,Gu:2018kmv}.
The charge profile of the particle content are shown in table \ref{modelt}.
It is to be noted that these additional fermions are vector-like, and hence, they do not introduce any new anomalies in theory ~\cite{ano1,ano2}.
The chiral gauge anomaly free condition coming from the one loop triple gauge boson vertex, which
reads \cite{Pal:1690642}:
\begin{equation}
	\sum_{rep}= Tr[\{T^a_L,T^b_L\}T^c_L]-Tr[\{T^a_R,T^b_R\}T^c_R]=0. \label{vl}
\end{equation}
Here, $T$ denotes the generators for the SM gauge group and $L, R$ denotes the interactions of
left or right chiral fermions with the gauge bosons. From Eq. \eqref{vl} it is clear that the SM satisfies the anomaly free condition because of the presence of a quark family to
each lepton family  \cite{Pal:1690642, Kannike:2016fmd}. On the other hand, the additional vector-like fermions introduced here, have the left chiral components transforming similarly to the right chiral ones under the SM gauge
symmetry. Therefore, the model is anomaly free.

All the BSM particles are considered odd under the discrete $Z_2$  symmetry, such that this BSM field does not mix with the SM fields. As a result, the lightest and neutral particle is stable and considered to be a viable dark matter candidate. Let us now elaborated on the model part in detail. The Lagrangian of the model read as,
\begin{equation}
\mathcal{L}=\mathcal{L_{\rm SM}}+\mathcal{L_S}+\mathcal{L_F}+\mathcal{L}_{int},
\end{equation}
where,
\begin{eqnarray}
\mathcal{L_S}&=&\frac{1}{2}|\partial_{\mu}S|^2-\frac{1}{2}kS^2\phi^2-\frac{1}{4}m_S^2S^2-\frac{\lambda_S}{4!}S^4\label{eq:scalar},\\
\nn \mathcal{L_F}&=&\overline{F}_D\gamma^{\mu}D_{\mu}F_D+\overline{E}_S\gamma^{\mu}D_{\mu} E_S-M_{ND}\overline{F}_DF_D - M_{NS}\overline{E}_SE_S,\\
\mathcal{L}_{int}&=&-Y_{N}\overline{F}_D\phi E_S - Y_{fi} \overline{L}_i F_D S + h.c.~,\label{lint}
\end{eqnarray}
$D_{\mu}$ stands for the corresponding covariant derivative of the doublet and singlet fermions. The SM Higgs potential is given by,  $V^{SM}(\phi)=-m^2\phi^2+\lambda\phi^4$, with,  
$\phi=( G^+, \frac{H+v+iG}{\sqrt{2}})^T$  is the SM Higgs doublet. $G$'s stand for the Goldstone bosons and $v=246.221$ GeV being the vacuum expectation value of the Higgs $H$ fields. The mass matrix for these charged fermion fields is given by,
\begin{eqnarray}
\mathcal{M}=\begin{pmatrix}
M_{ND}&M_X\\M_X^{\dagger}&M_{NS}\\
\end{pmatrix},
\label{eq:mass}
\end{eqnarray}
where, $M_X=\frac{Y_{N}v}{\sqrt{2}}$. The charged component of the fermion doublet ($E_D^\pm$) and the singlet charged fermion ($E_S^\pm$) mix at tree level. 
The mass eigenstates are obtained by diagonalizing the mass matrix with
a rotation of the ($E_D^\pm$  $E_S^\pm$) basis,
\begin{eqnarray}
\begin{pmatrix}
E_1^\pm\\E_2^\pm\\
\end{pmatrix}=\begin{pmatrix}
\cos\beta&\sin\beta\\-\sin\beta&\cos\beta\\
\end{pmatrix}\begin{pmatrix}
E_D^\pm\\E_S^\pm\\
\end{pmatrix}.
\end{eqnarray}
The mixing angle $\beta$ between the fermions can be written as,
\begin{equation}
\tan 2 \beta = \frac{2 M_X}{M_{NS}-M_{ND}}.\nn
\end{equation}
Diagonalization of eqn.~\ref{eq:mass} gives the following eigenvalues for the charged leptons ($M_{NS}-M_{ND} \gg M_X$) as,
\begin{eqnarray}
M_{E_1^\pm} &=& M_{ND} - \frac{2 (M_X)^2}{M_{NS}-M_{ND}}, \nn\\
M_{E_2^\pm} &=& M_{NS} + \frac{2 (M_X)^2}{M_{NS}-M_{ND}}.\nn
\end{eqnarray}
The masses of the neutral fermion scalar fields can be calculated as,
\begin{eqnarray}
M_{X_1^0}=M_{ND}, \, M_S^2=\frac{m_S^2+kv^2}{2}~ \ \text{and}~\ M_H^2= 2\lambda v^2. \nn
\end{eqnarray} 
Hence, in this model, neutral fermion can not be the DM candidate as $M_{E_1^\pm}<M_{X_1^0}<M_{E_2^\pm}$. Only the scalar fields $S$ for $M_S < M_{E_1^\pm}$ can behave as a viable DM candidate. We keep $M_{E_2^\pm}=1500$ GeV and $\cos\beta=0.995$ fixed through out the analysis. We will provide a detailed discussion on the new region of the allowed parameter spaces and the effect of the presence of additional $Z_2$-odd fermion in the dark matter section~\ref{dm1}.

The parameter space of this model is constrained by various bounds arising from theoretical considerations like absolute vacuum stability and unitarity of the scattering matrix, observation phenomenons like dark matter relic density. The LHC also puts severe constraints on this model. In the following section, we discuss constraints associated with the model.
\section{Constraints on this models}\label{s3}
Scotogenic model parameter space is constrained from theoretical considerations like  absolute vacuum stability, perturbativity and unitarity of the scattering matrix.  The direct search limits at LEP and electroweak precision measurements put severe restrictions on the model.
The recent measurements of the Higgs invisible decay width and signal strength at the LHC put additional constraints.
The requirement that the dark matter (DM) saturates the DM relic density all alone restricts the allowed parameter space considerably. Although some of these constraints are already discussed in the literature. We discuss a few constraints considered in our model in the following subsections.

\subsection{Constraints on scalar potential couplings from stability, perturbativity and unitarity} 
Most severe constraints come from the `bounded from below' of the potential, which ensures the absolute stability of the electroweak vacuum. The potential bounded from below signifies that there is no direction in field space along which the potential tends to minus infinity. In unitary gauge, for $H,S>>v$, the scalar potential of equation~(\ref{eq:scalar}) can be further simplified as,
\begin{eqnarray}
\nn V(H,~S) = \frac{1}{4}\left\lbrace \sqrt{\lambda} H^2+ \sqrt{\frac{\lambda_S}{6}} S^2 \right\rbrace^2 + \frac{1}{4}\left\lbrace\kappa +   \sqrt{\frac{2 \lambda \lambda_S}{3}}\right\rbrace H^2 S^2.
\label{scalpotstability}
\end{eqnarray}
The necessary conditions  for the scalar potential are given by,
\be
\lambda(\Lambda) > 0, \quad \lambda_S(\Lambda) > 0 \quad {\rm and} \quad \kappa(\Lambda) + \sqrt{\frac{2 \lambda(\Lambda) \lambda_S(\Lambda)}{3}} > 0.
\nonumber
\ee
Here, all the coupling constants in this model are evaluated at a scale $\Lambda$ using RG equations \cite{Garg:2017iva}. 
However, these conditions become non-functional if the Higgs quartic coupling $\lambda$ becomes negative at some energy scale to contribute to the electroweak vacuum metastable. In this situation, we need to handle metastability constraints on the potential difference, shown in Ref.~\cite{Khan:2012zw}.
Besides, for the radiatively improved Lagrangian of our model to be perturbative, we have~\cite{Lee:1977eg, Cynolter:2004cq},
\be
\lambda (\Lambda)\, 
 <\, \frac{4\pi}{3} \, ; \,\,\, |\kappa(\Lambda)|\, < \,8\pi \,;\,\,\, |\lambda_{S}(\Lambda)|\, < \, 8\pi.
 \ee
The couplings of the scalar potential ($\lambda,\kappa$ and $\lambda_S$) of this model are constrained by the unitarity of the scattering matrix (S-matrix). At very high field values, one can obtain the S-matrix by using various scalar-scalar, gauge boson-gauge boson, and scalar-gauge boson scatterings. Using the equivalence theorem, we reproduced the S-matrix for this model. The unitarity demands that the eigenvalues of the S-matrix should be less than $8\pi$. The unitary bounds are given by~\cite{Cynolter:2004cq}, 
\bea
\lambda \leq 8 \pi ~{\rm and}~ \Big| 12 {\lambda}+{\lambda_S} \pm \sqrt{16 \kappa^2+(-12 {\lambda}+{\lambda_S})^2}\Big| \leq 32 \pi.\nonumber
\eea
\subsection{LHC diphoton signal strength bounds}
At one-loop level, the physical charged fermion $E_1^{\pm}$ and $E_2^{\pm}$ add extra contribution to the decay width as,
\begin{eqnarray}
\Gamma(H\rightarrow \gamma\gamma)=A\Big| \sum_i Q_i^2 Y_{Ni} F_{1/2}(\tau_{E_i^\pm}) + C \Big|,\,\,\,\,
\end{eqnarray}
where, $A=\frac{\alpha^2M_h^3}{256\pi^3v^2}$, $C$ is the SM contribution, $ C=\sum_fN_f^cQ_f^2y_fF_{1/2}(\tau_{E_i^{\pm}})+y_WF_1(\tau_W)$ and $\tau_x=\frac{M_H^2}{ M_X^2}$. $Q$ denote electric charge of corresponding particles and $N_f^c$ is the color factor. Higgs $H$ coupling to $f\overline{f}$ and $WW$ is denoted by $y_f$ and $y_W$. $Y_{N1}= \sqrt{2} \cos\beta \sin\beta Y_{N}$ and $Y_{N2}= -\sqrt{2} \cos\beta \sin\beta Y_{N}$ stand for corresponding couplings $H \, E_i+E_i^-$ ($i=1,2$) and the loop function $F_{(0,1/2,1)}(\tau)$ can be found in Ref~\cite{Djouadi:2005gj}.
In this analysis, we find that $M_{E_{1,2}^\pm}>200$ GeV for $Y_N=\mathcal{O}(1)$ is still allowed from the LHC di-photon signal strength $\mu_{\gamma\gamma}=\frac{\Gamma(H\rightarrow\gamma\gamma)_{BSM}}{\Gamma(H \rightarrow\gamma\gamma)_{SM}}$ data. 
\subsection{Bounds from electroweak precision experiments}
Bounds from electroweak precision experiments are added in new physics contributions via self-energy parameters $S,T,U$ from EW precision experiments does put bounds on new physics contributions~\cite{Baak:2014ora, Peskin:1991sw}.
The $S$ and $T$ parameters allow the new physics contributions to the neutral and the difference between neutral and charged weak currents, respectively. However, the $U$ parameter is only sensitive to the mass and width of the $W$-boson. Thus in some cases, this parameter is neglected. 
The NNLO global electroweak fit results from the Gfitter
group~\cite{Baak:2014ora} gives, $\Delta S_{BSM}<0.05\pm0.11$, $T_{BSM}<0.09\pm0.13$ and $\Delta U_{BSM}<0.011\pm0.11$.
In this model, a tiny mass difference $\Delta M \sim 20$ GeV between the charged and neutral fermions of the doublet $F_D$~\cite{Peskin:1991sw, Cynolter:2008ea} with $M_{ND}>200$ GeV and heavy singlet charged fermion mass $\mathcal{O}(1)$ TeV are considered to evade these bounds. We also keep fixed $\cos\beta=0.995$ and $M_{E_2^\pm}=1500$ GeV throughout the analysis to avoid the bounds from the Electroweak Precision Parameters.

\subsection{Dark matter}
The lightest stable $Z_2$ odd particle, $S$ behaves like a proper DM candidate in our model. As per our choice of parameter space, DM relic density constraints should satisfy current results from Planck and WMAP~\cite{Aghanim:2018eyx},
\be \Omega_{DM}h^2  \, = \,  0.1198 \, \pm \, 0.0012 .\ee
Recent direct-detection experiments like the Xenon-1T~\cite{Aprile:2018dbl} and invisible Higgs decay width data including indirect Fermi-LAT data~\cite{fermilat1} have restricted the arbitrary Higgs portal coupling and the dark matter mass~\cite{Khan:2012zw,Athron:2017kgt,indirect2}. It is also possible to explain various observations in the indirect DM detection experiments from this model. However, we do not discuss these here, as these estimations involve proper knowledge of the astrophysical backgrounds and an assumption of the DM halo profile, which contains some arbitrariness.

In our study, we use {\tt FeynRules}~\cite{Alloul:2013bka} along with {\tt micrOMEGAs}~\cite{Belanger:2018mqt} to compute the relic density of the scalar DM. We present a comprehensive discussion on dark matter in the numerical analysis section. 
\subsection{Lepton flavour violation ($\mu\rightarrow e\gamma$) and anomolus magnetic moment}
\begin{figure}[h]
	\centering
		\includegraphics[scale=.95]{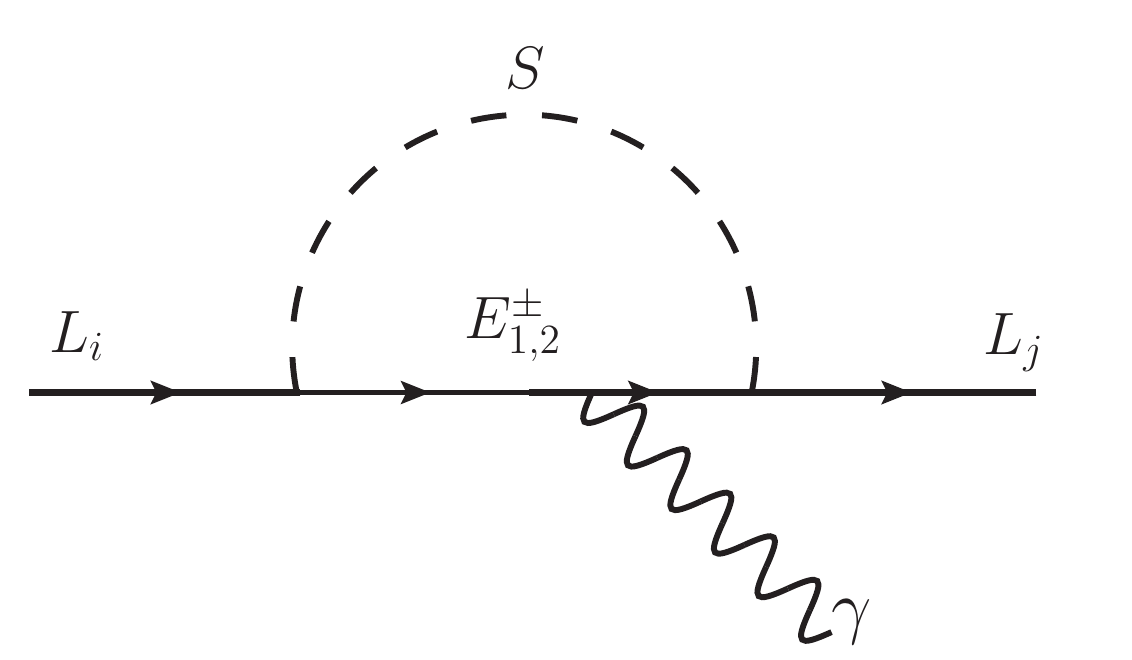}
		\caption{ Muon and electron anomolus magnetic moment and LFV process $\mu\rightarrow e\gamma$ decay diagrams mediated by charged particles $E_1^{\pm}$ and $E_2^{\pm}$. }\label{lfv}
\end{figure}

It is a well-known lepton flavour violation (LFV) process that put severe constraints on the LFV couplings and, in general, on the model parameter space. The size of the LFV is controlled by the lepton number violating couplings $Y_{fi} ~(i=1,2,3)$. Since the observed dark matter abundance is typically obtained for $\kappa=\mathcal{O}(0-1)$ and $Y_{fi}=\mathcal{O}(0-1)$ through $s$-channel,  $t$-channel annihilation and the combination of these two processes (co-annihilation, i.e., mass differences can also play a crucial role). The lepton flavour observables are expected to give additional stringent constraints on the parameter spaces. 
Among the various LFV processes, the radiative muon decay
$\Gamma(\mu\rightarrow e \gamma)$ is one of the popular and restrictive one, which in the present model is mediated by charged particles $E_1^\pm, E_2^\pm$ present in the internal lines of the
one-loop diagram~\ref{lfv}. 
The corresponding expression for the branching ratio is given by,
\begin{eqnarray}
 {\rm BR} (\mu\rightarrow e \gamma) = \frac{3 \alpha_{em}}{64 \pi G_F^2} \Big | \cos^2\beta \, Y^\dagger_{f1} Y_{f2} \frac{ F(M_{E_{1}^\pm}^2/M_S^2) }{ M_S^2}+ \sin^2\beta \, Y^\dagger_{f1} Y_{f2} \frac{ F(M_{E_{2}^\pm}^2/M_S^2) }{ M_S^2}\Big |^2 ,
\label{eq:fl}
\end{eqnarray}
where, $
F(x)=\frac{x^3-6 x^2 +3 x+ 2 + + x ln(x)}{6 \, (x-1)^4}
$.
The most recent experimental bounds for LFV could be found in Ref.~\cite{Baldini:2018nnn}. Throughout this analysis we keep fixed $Y_{f2}=\mathcal{O}(10^{-3})$ and put constraints to the other parameters from the flavour violating decay~\cite{Baldini:2018nnn} $ {\rm BR} (\mu\rightarrow e \gamma) < 4.2 \times 10^{-13}$ at $90\%$ CL.

Due to the presence of vector-like fermion, the new contribution to anomalous magnetic moment can be written as~\cite{Harnik:2012pb},
\begin{eqnarray}
 \Delta \alpha_{i}&=\frac{ m_i \, m_{E_1^{\pm} } \, A_{f2}^2 }{ 256\, \pi^2 \, m_S^2} \Big( 2 \, \text{ln}\, \frac{m_s^2 }{ m_{E_1^{\pm}}^2} -3  \Big) + \frac{ m_i \, m_{E_2^{\pm} } \, B_{f2}^2 }{ 256\, \pi^2 \, m_S^2} \Big( 2 \, \text{ln}\, \frac{m_s^2 }{ m_{E_2^{\pm}}^2} -3  \Big) 
\end{eqnarray}
where $A_{fi}=Y_{fi}\cos\beta$ and $B_{fi}=Y_{fi}\sin\beta$.
The discrepancy between the theoretical SM predictions and the experimental values are given by~\cite{Bennett:2006fi, Parker:2018vye}:
$\delta a_{\mu}=a_{\mu}^\text{exp}-a_{\mu}^\text{SM}=(2.74\pm 0.73)\times 10^{-9}$
and 
$\delta a_{e}=a_{e}^\text{exp}-a_{e}^\text{SM}=-(8.8\pm 3.6)\times 10^{-13}$.
With the choices of appropriate parameters in this model, we can explain the electron anomalous magnetic moment, still not the muon anomalous magnetic moment (we have $\delta a_{\mu}\sim 10^{-14}$) at the same time. The parameters which satisfy the discrepancy of muon anomalous magnetic moment violates the LFV data. We will not focus on this further.
\subsection{Neutrino mass via one loop process}\label{s4}
\begin{figure}[h]
\centering
	\includegraphics[scale=.95]{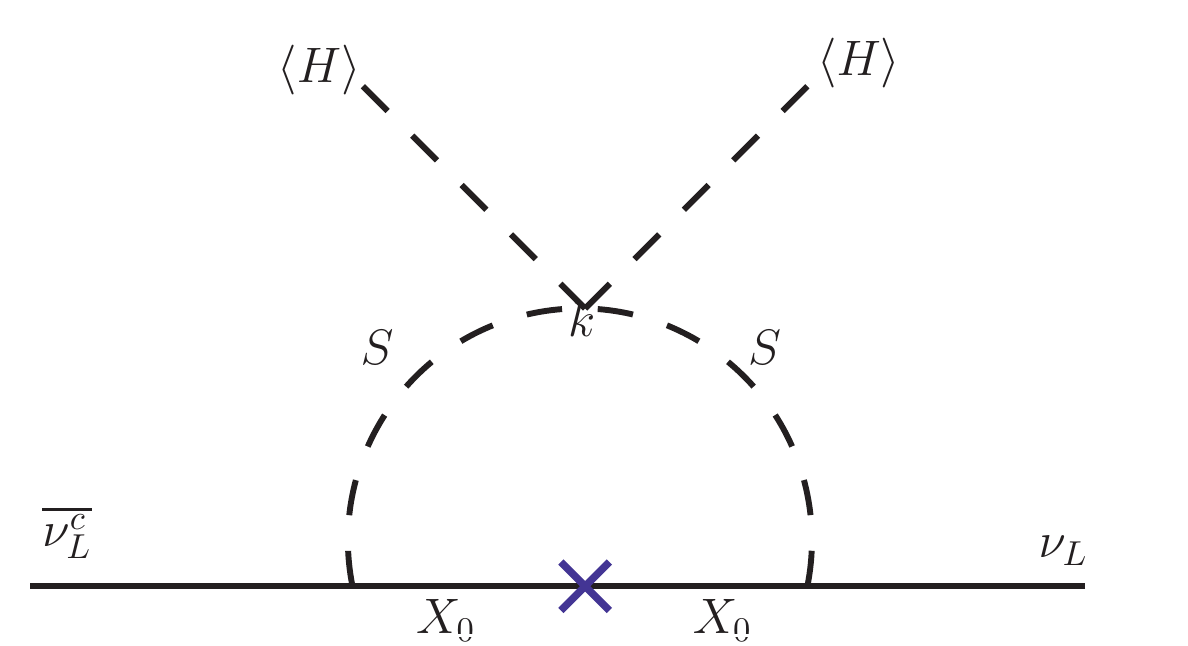}
	\caption{ \rm One-loop contribution to neutrino mass generation with a scalar $S$ and fermion $X_1^0$.}\label{numass}
\end{figure}

In this section, we will try to give a brief overview of the neutrino mass generation at the one-loop level. 
The neutral $Z_2$-odd scalar and fermion involved in the radiative neutrino mass generation after the EWSB, which is shown in Fig.~\ref{numass}. 
Summing over all the two-point function contributions, we arrive at the neutrino mass matrix component as~\cite{FileviezPerez:2009ud},
\begin{eqnarray}
(M_{\nu})_{ij}=\frac{1}{16\pi^2} (Y_{fi}^\dagger Y_{fi}) ( \, \kappa v^2 ) I (M_N, M_{DM}),  
\label{eq:n1}
\end{eqnarray}
where, $i,j=1,2,3$ stand for the lepton generation index. $M_N$ is the mass for the neutral heavy fermion. $I (M_N, M_{DM}) $ is the loop function, defined as~\cite{FileviezPerez:2009ud},

\begin{eqnarray}
I (M_N, M_{DM}) = 4 M_N  \frac{M_{DM}^2 -M_N^2 + M_N \, log(\frac{M_N^2}{M_{DM}^2})}{(M_{DM}^2-M_N^2)^2}.
\label{eq:n2}
\end{eqnarray}
To get the neutrino mass eigenvalues, we have to diagonalize the above mass matrix using the well established PMNS matrix as: $ m_{Diag}=U_{\rm PMNS}^\dagger M_\nu U_{\rm PMNS}$. It is also essential to ensure that the choice of Yukawa couplings, as well as other parameters
involved in light neutrino mass, are consistent with the current neutrino oscillation data.

From the above equations~\ref{eq:n1} and~\ref{eq:n2}, light neutrino masses, and mixing angles can be visualized by adjusting the coupling and mass parameters present in equation \eqref{eq:n1}. For a few hundred GeV dark matter and heavy neutral fermions, one can choose small $\kappa$ of $\mathcal{O}(10^{-6})$ to get the small neutrino masses. From Eq. \eqref{eq:n1} it is clear that, in the limit $\kappa v^2\rightarrow0$, light neutrino mass vanishes. This limit also signifies the fact that the vanishing neutrino masses are quite obvious as $\kappa$ in the scalar potential breaks lepton number by two units, when considered together with the SM-singlet fermions Lagrangian. Hence, the smallness of $\kappa$ is technically natural in the 't Hooft sense~\cite{tHooft:1980xss}, as adjusting $\kappa\rightarrow0$ allows us to define global $U(1)$ lepton number symmetry. 
At the same time, by adjusting both the real and imaginary parts of the Yukawa couplings, the mixing angles could be produced. This smallness of the Higgs portal coupling enhances the allowed region of the parameter space, and the relic density could produce via the other channels, which we will discuss in detail in the dark matter numerical analysis section. The analysis of neutrino mass carried out in this work is more of a perfunctory rather than being comprehensive.
\section{Numerical analysis}\label{s5}
\subsection{Dark matter}\label{dm1}
 \begin{figure}[h!]
	\begin{center}
		\subfigure[]{
			\includegraphics[scale=0.6]{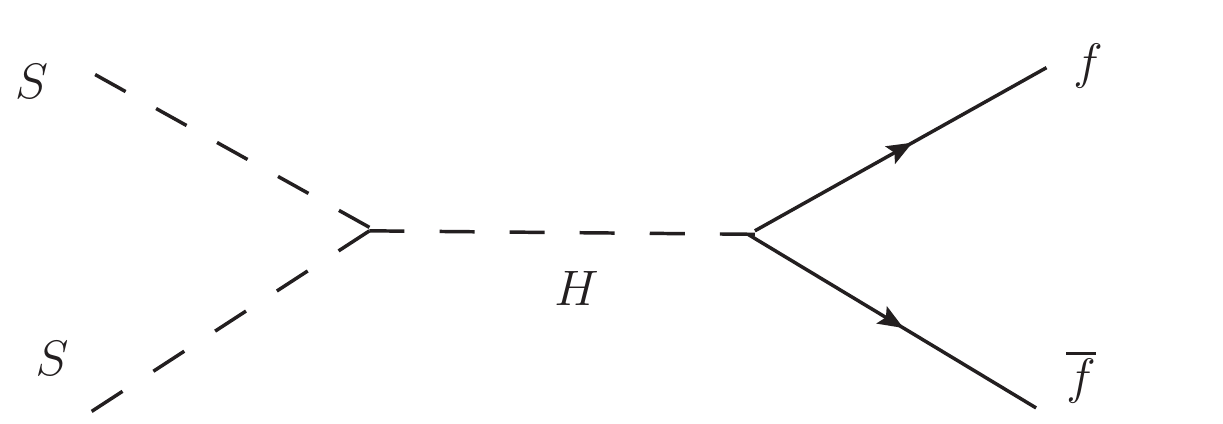}}
		\hskip 1pt
		\subfigure[]{
			\includegraphics[scale=0.6]{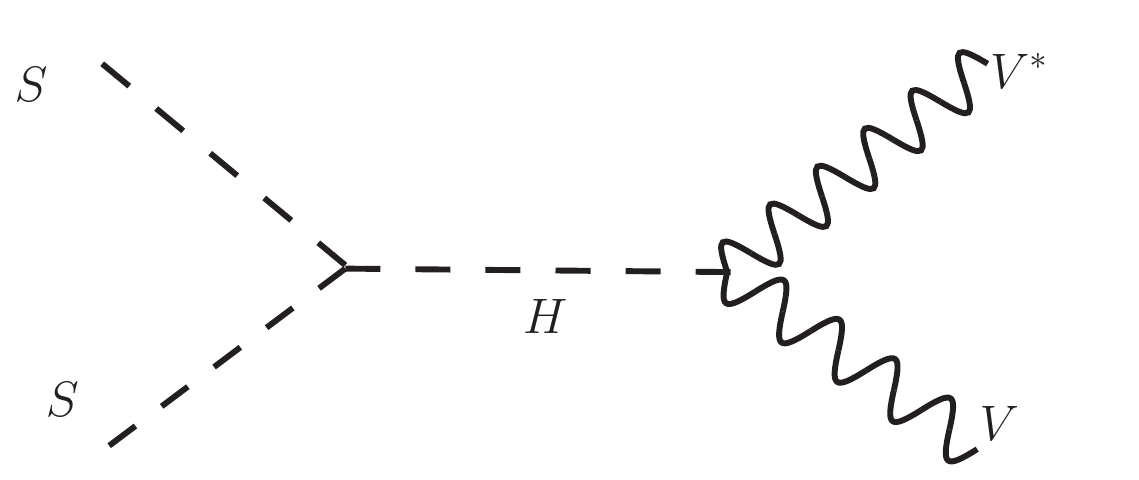}}
		\subfigure[]{
			\includegraphics[scale=0.6]{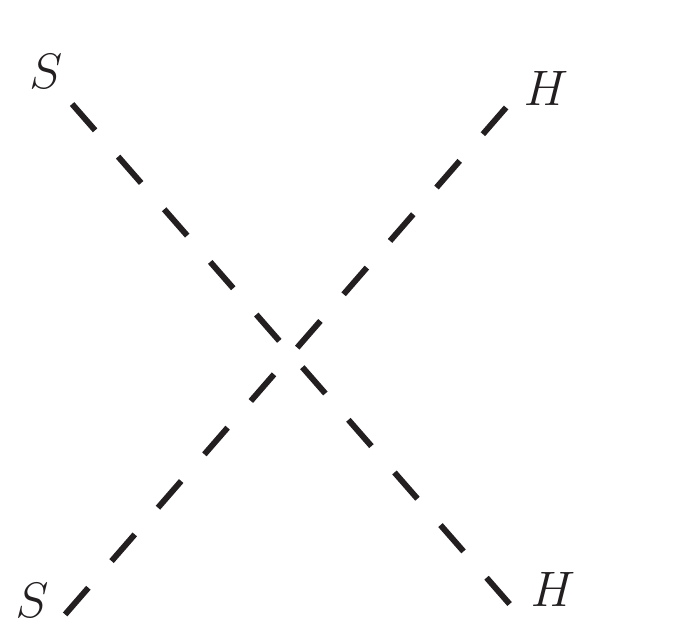}}
		\hskip 1pt
		\subfigure[]{
			\includegraphics[scale=0.6]{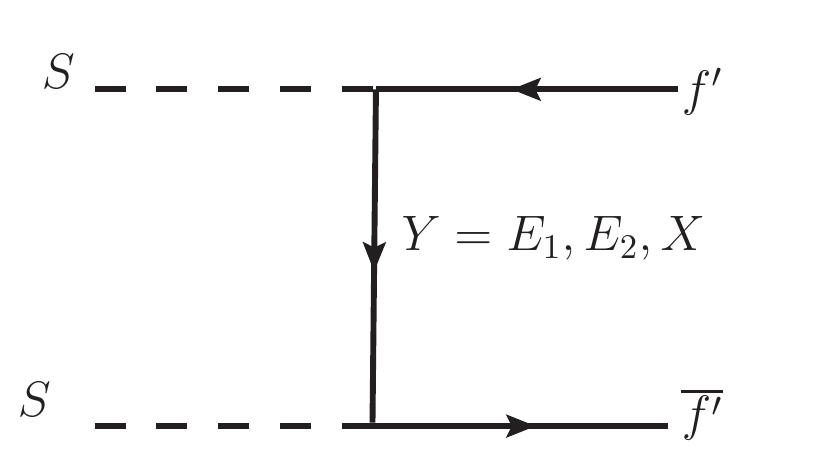}}
		\caption{ \it The DM annihilation diagrams give the relic density. $V$ stands for gauge bosons $W,Z$, $f'$ represents the SM leptons and $f$ are SM leptons and quarks.}
		\label{fig:DarkAn}
	\end{center}
\end{figure}
\begin{figure}[h!]
	\begin{center}
		\subfigure[]{
			\includegraphics[scale=0.6]{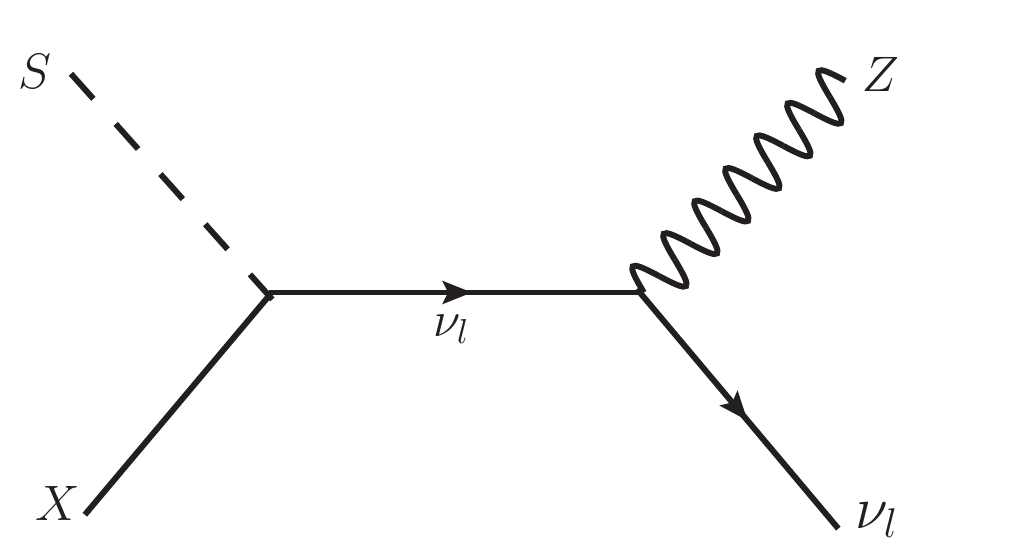}}
		\hskip 1pt
		\subfigure[]{
			\includegraphics[scale=0.6]{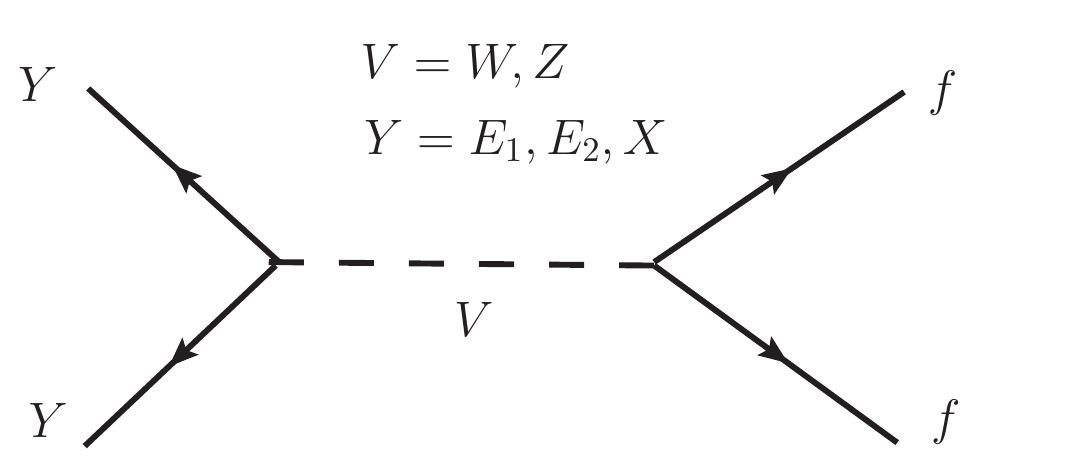}}
		\subfigure[]{
			\includegraphics[scale=0.6]{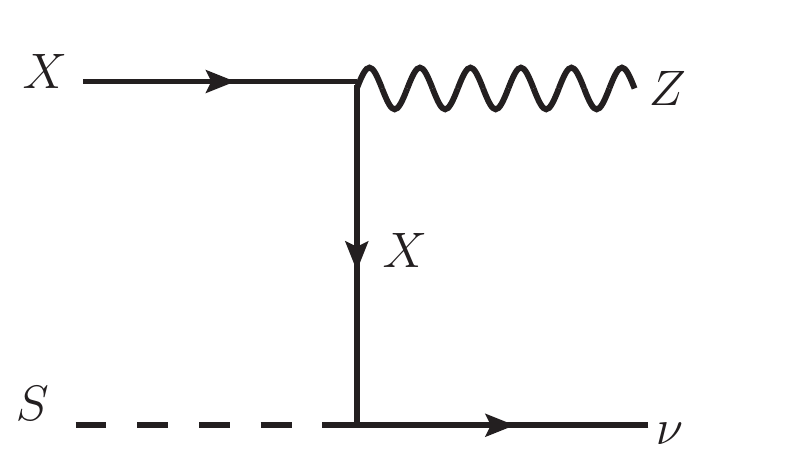}}
		\hskip 1pt
		\subfigure[]{
			\includegraphics[scale=0.6]{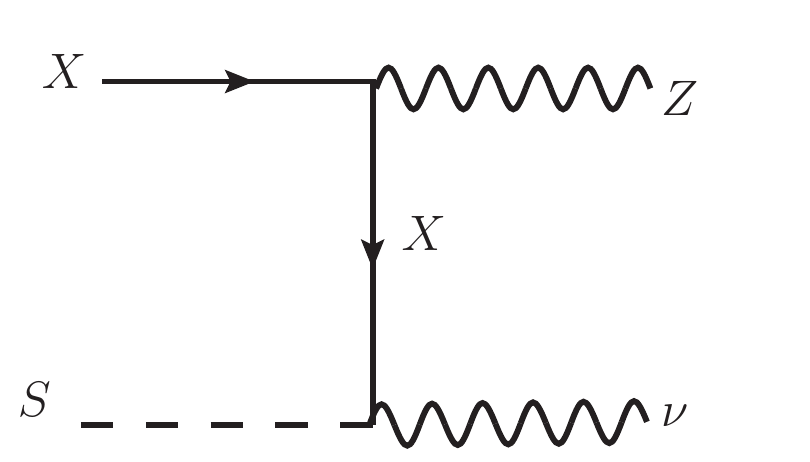}}
		\caption{ \it The Co-annihilation and annihilation diagrams of the DM and the other $Z_2$-odd fermion fields. $f$ are SM leptons and quarks.
		}
		\label{fig:DarkCoan}
	\end{center}
\end{figure}
As pointed out in the previous section, the viable DM candidate in this model is the lightest $Z_2$-odd singlet scalar $S$. The production mechanism of this DM  candidate  depends upon the Higgs portal couplings $\kappa$ through $s$- and cross-channels (see Figs.~\ref{fig:DarkAn}-(a), ~\ref{fig:DarkAn}-(b) and ~\ref{fig:DarkAn}-(c)).
\begin{figure}[h!]
	\begin{center}{
			\includegraphics[scale=0.43]{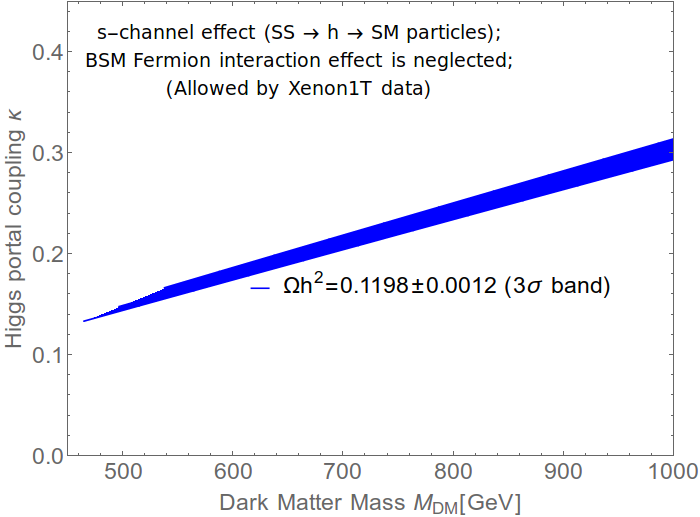}}
		\caption{ \it  The dark matter relic density through $s$- and cross- channels only, with direct detection and other theoretical and experimental constraints. The $3\sigma$  relic density $\Omega h^2=0.1198\pm 0.0012$ constraint is shown
			as the blue band. The Yukawa couplings $Y_{fi}$ and $Y_N$ are taken to be zero.}
		\label{fig:DarkAnS}
	\end{center}
\end{figure}
 \begin{figure}[h!]
	\begin{center}{
			\includegraphics[scale=0.43]{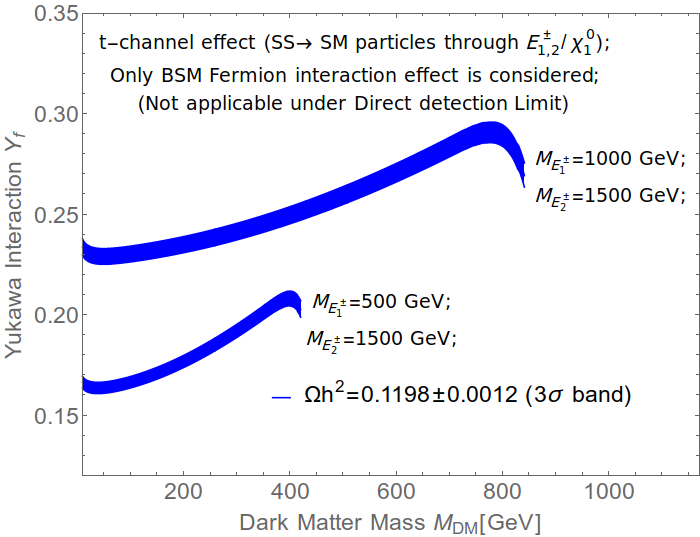}}
		\caption{ \it  The dark matter relic density through $t$- channels only, with direct detection and other theoretical and experimental constraints. The $3\sigma$  relic density $\Omega h^2=0.1198\pm 0.0012$ constraint is shown
as the blue band. The Higgs portal couplings $\kappa$ is taken to be zero.}
		\label{fig:DarkAnT}
	\end{center}
\end{figure}
It is to be noted that in presence of the Yukawa couplings $Y_{fi}$ and $Y_N$, a huge improvement to the region of the dark matter parameter space is noticed here. Depending upon the size of the Yukawa couplings $Y_{fi}$, one can get a dominant DM annihilation through $t$- and $u$-channels (see Fig.~\ref{fig:DarkAn}-(d)) in our model.
The interference between the $s$-channel, cross-channel and $t,u$-channels also played a crucial role to achieve the correct DM density\footnote{It is to be noted that the Sommerfeld enhancement don't play any role to enhance the current dark matter phenomenology \cite{ArkaniHamed:2008qn} and $M_{E_{1,2}^{\pm}}>M_{DM}$.}. The co-annihilation channels (e.g., see Fig.~\ref{fig:DarkCoan}) also played an important role in getting a viable region of allowed dark matter parameter space.
\begin{table*}[h!]
	\centering
	\begin{tabular}{|p{1.6cm}|p{1.2cm}|p{1.1cm}|p{1.2cm}|p{1.2cm}|c|p{4.7cm}|}
	\hline
	\hline
	Channel & $M_{DM}$ (GeV) &~~ $\kappa~~~$& $M_{E_1^\pm}$ (GeV) &$Y_{f}$&$\Omega_{DM}h^2$&~~~~~~~~~~Percentage \\
	\hline
		&&&&&&$\sigma(S S\rightarrow W^\pm W^\mp)~~~47\%$ \\
		~~BP-a1&570&0.1703&2000~~&0.0&0.1198&$\sigma(S S\rightarrow HH)\quad~24\%$ \\
		&&&&&&$\sigma(S S\rightarrow ZZ)\quad23\%$\\
		&&&&&&$\sigma(S S\rightarrow t\bar{t})\quad6\%$\\
		\hline
		&&&&&&$\sigma(S S\rightarrow \nu \nu)\quad~98\%$\\
	~~BP-b1&10&0.0&500~~&0.1665&0.1198& $\sigma(SS \rightarrow  ll)\quad 2 \%$\\
		\hline
		&&&&&&$\sigma(S S\rightarrow \nu \nu)\quad~98\%$\\
	~~BP-b2&60&0.0&500~~&0.1640&0.1198& $\sigma(SS \rightarrow  ll)\quad 2 \%$\\
\hline
	&&&&&&$\sigma(S S\rightarrow \nu \nu)\quad~98\%$\\
	~~BP-b3&100&0.0&500~~&0.1677&0.1198& $\sigma(SS \rightarrow  ll)\quad 2 \%$\\
\hline
\hline
	\end{tabular}
	\caption{The benchmark points allowed by all the theoretical and experimental constraints. The density of the dark matter $S$ is dominated by either $s$- or $t,u$-channel annihilation processes. We consider $Y_{f1}=Y_{f3}=Y_f$ to avoid flavour violating decay processes.}
	\label{tabDM:1}
\end{table*}
 \begin{figure}[h!]
	\begin{center}{
			\includegraphics[scale=0.43]{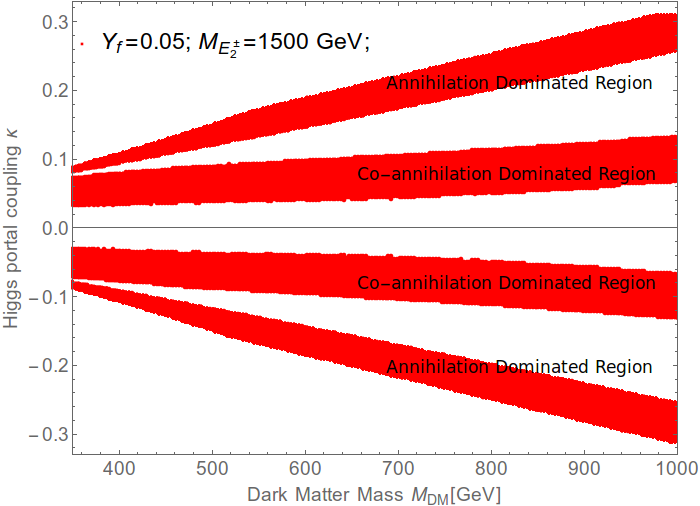}}
		\caption{ \it  The coupling $y_f=0.05$ and  second charged fermion mass  $M_{E_2^\pm}=1500$ GeV are fixed. $M_{DM}$, $\kappa$ and $M_{E_1^\pm}$ parameters are varied in this plot. These red points satisfy the relic density at $3\sigma$ C.L. with $\Omega h^2=0.1198$ $\pm0.0012$, satisfying all the theoretical and experimental bounds.} \label{fig:DarkAnr}
	\end{center}
\end{figure}
\begin{figure}[h!]
	\begin{center}{
			\includegraphics[scale=0.43]{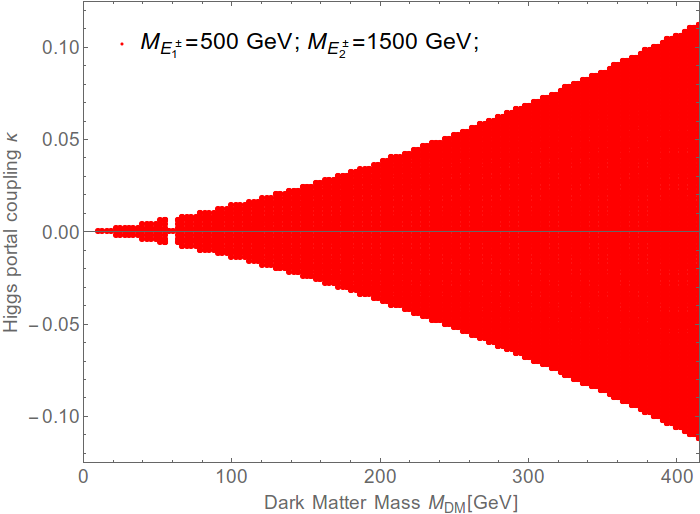}}
		\caption{ \it  The first and  and  second charged fermion masses $M_{E_1^\pm}=500$ GeV and  $M_{E_2^\pm}=1500$ GeV are fixed. $M_{DM}$, $\kappa$ and $y_f$ parameters are varied in this plot. These red points satisfy the relic density at $3\sigma$ of $\Omega h^2=0.1198$ $\pm0.0012$ and pass all the theoretical and experimental bounds.}
		\label{fig:DarkAll1}
	\end{center}
\end{figure}
\begin{figure}[h!]
	\begin{center}{
			\includegraphics[scale=0.43]{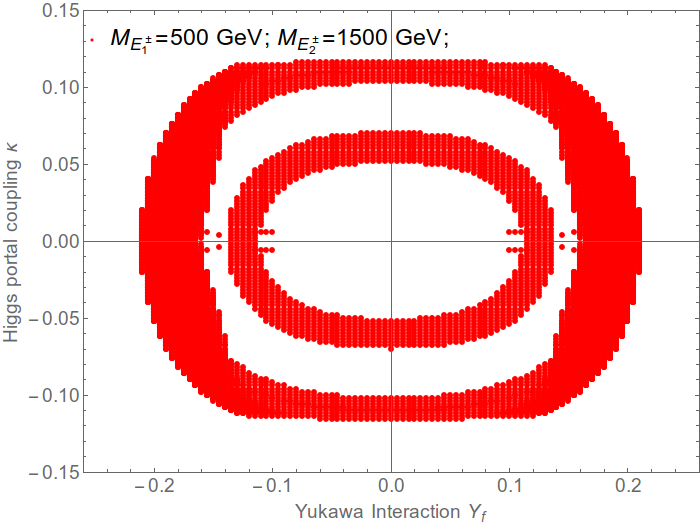}}
		\caption{ \it  The first and  and  second charged fermion masses $M_{E_1^\pm}=500$ GeV and  $M_{E_2^\pm}=1500$ GeV are fixed. $M_{DM}$, $\kappa$ and $y_f$ parameters are varied in this plot. These red points satisfy the relic density at $3\sigma$ of $\Omega h^2=0.1198$ $\pm0.0012$ and pass all the theoretical and experimental bounds. The core one gives the overabundant relic density on the other hand, the outer ring gives the under-abundent relic density for the given dark matter mass. Here, for large $\kappa$, it will violate the direct detection cross section.  It is to be noted that, the interference effect of the s-channel ($\kappa$) and t-channel ($Y_f$) will play important role in relic density calculation.}
		\label{fig:DarkAll2}
	\end{center}
\end{figure}

It is already evident that if we neglect the effect of other $Z_2$-odd fermions, {\it i.e.}, annihilation through $t$-channels and other co-annihilation processes, a very small $low$-DM mass region around $55{\text{GeV}}<M_{DM}<70$GeV for Higgs portal coupling $\kappa\sim 0.005$ is giving the exact relic density, allowed by the direct detection~\cite{Aprile:2018dbl} and LHC data. The main dominant channels for $low$-DM mass region is $SS \rightarrow b \bar{b}$.
For $M_{DM} > 100$ GeV, $SS \rightarrow V V$, where $V=W^\pm,Z$ gauge bosons~\cite{McDonald:1993ex} dominates over other DM annihilation channels. Under the approximation
$M_{DM} >> M_V, M_H$ , in the non-relativistic limit one can get the DM annihilation cross-section as $\sigma (SS \rightarrow W^+ W^-) \propto \frac{k^2}{M_{DM}^2}$. The allowed relic density (dominated by $s$- and cross-channels only) for the $high$-DM mass region in $\kappa-M_{DM}$ plane is displayed in Fig.~\ref{fig:DarkAnS}. We also present corresponding benchmark points BP-1a and the percentage of different annihilation channel's contributions in the Tab.~\ref{tabDM:1}. As usual, the main dominant channels are $SS \rightarrow YY$ with $Y=W,Z$ and $H$ for the $high$-DM mass region.
The $3\sigma$  relic density $\Omega h^2=0.1198\pm 0.0012$ constraint is shown
as the blue band.  One can get the exact relic density for the DM-mass region 70 GeV$< M_{MD} <  450$ GeV, however, it is ruled out by the present direct detection cross-section~\cite{Aprile:2018dbl}.
So far, we do not have any direct signature of DM in the direct detection experiments, which suggest that we may have the dark matter with a $tiny$ or $zero$ Higgs portal coupling. Furthermore, the remaining effective cross-section $<\sigma_{eff} v>$ can be adjusted by the other annihilation and co-annihilation processes to achieve the exact dark matter density. In this model, we adopted such scenarios to achieve our goals.
For example, various dark matter masses can get the exact density with vanishing Higgs portal coupling ($\kappa$) by adjusting the charged fermion mass and Yukawa couplings $Y_{fi}$.
We portrait such variation in $Y_f-M_{DM}$ plane in Fig.~\ref{fig:DarkAnT} for two different values of charged fermion mass $M_{E_1^\pm}=500$ GeV and $M_{E_1^\pm}=1000$ GeV. We also consider $Y_{f1}=Y_{f3}=Y_f$ and $Y_{f2}=\mathcal{O}(10^{-3})$ to avoid the flavour violating decay processes (see eqn.~\ref{eq:fl}). 
\begin{figure}[h!]
	\begin{center}{
			\includegraphics[scale=0.43]{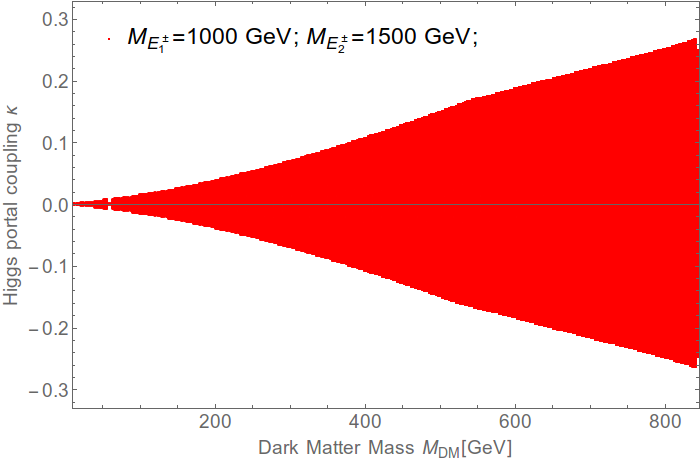}}
		\caption{ \it  The first and  and  second charged fermion masses $M_{E_1^\pm}=1000$ GeV and  $M_{E_2^\pm}=1500$ GeV are fixed. $M_{DM}$, $\kappa$ and $y_f$ parameters are varied in this plot. These red points satisfy the relic density at $3\sigma$ of $\Omega h^2=0.1198$ $\pm0.0012$ and pass all the theoretical and experimental bounds.}
		\label{fig:DarkAll3}
	\end{center}
\end{figure}
It can be noticed from Fig.~\ref{fig:DarkAnT} that one could get exact relic density for the dark matter mass as low as $M_{DM} = 10$ GeV.
As $\kappa=0$, the parameter space $M_{DM}<\frac{M_H}{2}$ is not restricted by the Higgs decay width and direct detection cross-section constraints. These data points also passed through other experimental constraints such as Higgs signal strength, electroweak precision test (EWPT) and theoretical bounds, viz., stability, unitarity, etc. The same $3\sigma$  relic density $\Omega h^2=0.1198\pm 0.0012$ constraint is shown
as the blue band.
The main dominant $t,u$-channel annihilation processes are $SS \rightarrow \nu \nu$ (see BP-b1,b2 and b3 in Tab.~\ref{tabDM:1}) and $SS \rightarrow l l$, where $l=e, ~\tau$  and $\nu=\nu_e,~\nu_\tau$ only as $Y_{f2}=\mathcal{O}(10^{-3})$. 

\begin{figure}[h!]
	\begin{center}{
			\includegraphics[scale=0.43]{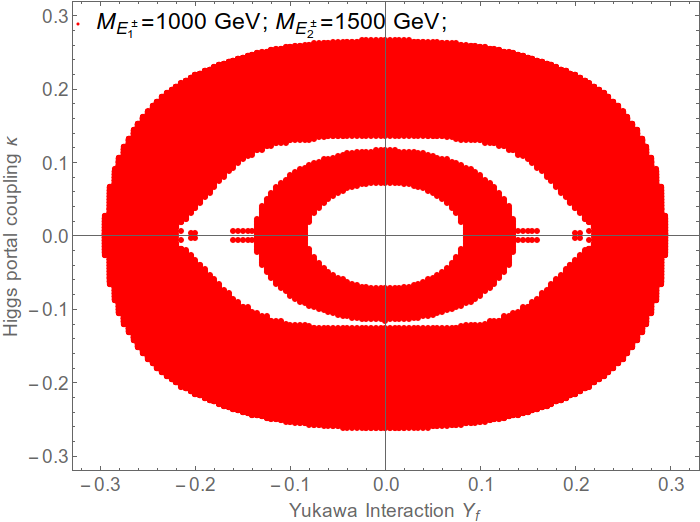}}
		\caption{ \it  The first and  and  second charged fermion masses $M_{E_1^\pm}=1000$ GeV and  $M_{E_2^\pm}=1500$ GeV are fixed. $M_{DM}$, $\kappa$ and $y_f$ parameters are varied in this plot. These red points satisfy the relic density at $3\sigma$ of $\Omega h^2=0.1198$ $\pm0.0012$ and pass all the theoretical and experimental bounds. The core one gives the overabundant relic density on the other hand, the outer ring gives the under-abundent relic density for the given dark matter mass. Here, for large $\kappa$, it will violate the direct detection cross section.  It is to be noted that, the interference effect of the s-channel ($\kappa$) and t-channel ($Y_f$) will play important role in relic density calculation.}
		\label{fig:DarkAll4}
	\end{center}
\end{figure}
\begin{table*}[h!]
	\centering
	\begin{tabular}{|p{1.6cm}|p{1.2cm}|p{1.3cm}|p{1.2cm}|p{1.2cm}|c|p{4.7cm}|}
		\hline
		\hline
		Channel & $M_{DM}$ (GeV) &~~~ $\kappa~~~$& $M_{E_1^\pm}$ (GeV) &$Y_{f}$&$\Omega_{DM}h^2$&~~~~~~~~~~Percentage \\
		\hline
		&&&&&&$\sigma(S S \rightarrow W^\pm W^\mp)\quad4\%$ \\
		~~BP-c1&501&-0.0384&582&-0.05&0.1233&$\sigma(S S \rightarrow HH)\quad 2\%$ \\
		&&&&&& $\sigma(S S \rightarrow ZZ)\quad2\%$ \\
		&&&&&& $\sigma(E_1^\pm E_1^\pm \rightarrow W^\pm W^\pm ) 65\%$  \\
		&&&&&&$\sigma(E_1^\pm E_1^\mp\rightarrow ZH)\quad20\%$\\
		&&&&&& $\sigma(E_1^\pm E_1^\pm \rightarrow  t\bar{t})\quad 3 \%$\\
		\hline
		&&&&&&$\sigma(S S \rightarrow W^\pm W^\mp)\quad19\%$ \\
		~~BP-c2&501&-0.087&586.2&-0.05&0.1162&$\sigma(S S \rightarrow HH)\quad 10\%$ \\
		&&&&&& $\sigma(S S \rightarrow ZZ)\quad9\%$ \\
		&&&&&& $\sigma(SS \rightarrow  t\bar{t})\quad 3 \%$\\
		&&&&&& $\sigma(E_1^\pm E_1^\pm \rightarrow W^\pm W^\pm ) 42\%$  \quad$\sigma(E_1^\pm E_1^\mp\rightarrow ZH)\quad13\%$	\\
		&&&&&& $\sigma(E_1^\pm E_1^\pm \rightarrow  t\bar{t})\quad 2 \%$\\
		\hline
		&&&&&&$\sigma(S S \rightarrow W^\pm W^\mp)\quad26\%$ \\ ~~BP-c3&501&-0.122&589.5&-0.05&0.1234&$\sigma(S S \rightarrow HH)\quad 15\%$ \\
		&&&&&& $\sigma(S S \rightarrow ZZ)\quad13\%$ \\
		&&&&&& $\sigma(SS \rightarrow  t\bar{t})\quad 4 \%$\\
		&&&&&& $\sigma(E_1^\pm E_1^\pm \rightarrow W^\pm W^\pm ) 30\%$  \quad$\sigma(E_1^\pm E_1^\mp\rightarrow ZH)\quad9\%$	\\
		&&&&&& $\sigma(E_1^\pm E_1^\pm \rightarrow  t\bar{t})\quad 1 \%$\\
		\hline
		&&&&&&$\sigma(S S \rightarrow W^\pm W^\mp)\quad44\%$ \\
		~~BP-c4&501&-0.148&595&-0.05&0.1166&$\sigma(S S \rightarrow HH)\quad 26\%$ \\
		&&&&&& $\sigma(S S \rightarrow ZZ)\quad22\%$ \\
		&&&&&&$\sigma(S S \rightarrow t\overline{t})\quad7\%$\\
		\hline
		\hline
	\end{tabular}
	\caption{The benchmark points allowed by all the theoretical and experimental constraints. The density of the dark matter $S$ is dominated by either annihilation or co-annihilation or combined effect of these processes.  We consider $Y_{f1}=Y_{f3}=Y_f$ to avoid flavour violating decay processes.}
	\label{tabDM:2}
\end{table*}
We now perform scans over the three dimensional parameter space. The mass parameter $M_{E_1^\pm}$ is varied from $200$ GeV (to avoid the experimental constraints) to $1000$ GeV with a step size $0.25$ GeV and $\kappa$ from $-0.35$ to $0.35$ with a step size $0.002$. The dark matter mass $M_{DM}$ from $\sim 200$ GeV to 1000 GeV with a step size $2$ GeV. 
The effect is almost negligible for the second charged fermion mass $M_{E_2^\pm}=1500$ GeV with $\cos\beta=0.995$. We fixed the coupling $Y_f$ at $0.05$ for the region $M_{E_1^\pm}>200$ GeV to reduce the  contributions through the $t,u$ annihilation channel in the relic density. 
For $\Delta M^{\pm,0}<0.1 M_{DM}$ \cite{Griest:1990kh} ($\Delta M^\pm= M_{E_1^\pm}-M_{DM}$ and $\Delta M^0= M_N-M_{DM}$), the co-annihilation channels (Fig.~\ref{fig:DarkCoan}) play an important role for the dark matter density calculation. 
In Fig.~\ref{fig:DarkAnr}, we display the allowed parameters in the $\kappa-M_{DM}$ plane.
These red points satisfy the relic density at $3\sigma$ C.L. with $\Omega h^2=0.1198$ $\pm0.0012$. The co-annihilation channels mainly dominate the two middle bands close to $\kappa \sim \pm0.03-\pm0.10$. For example, we present two such benchmark points (BP-c1 and BP-c2) and the corresponding contributions in Tab.~\ref{tabDM:2}. The other two bands in Fig.~\ref{fig:DarkAnr} are mainly dominated by the dark matter annihilation through $s+cross$-channels (see Fig.~\ref{fig:DarkAn}(a,b and c)). It also have small contribution from the dark matter annihilation through $t+u$-channels (see Fig.~\ref{fig:DarkAn}(d)). Large Higgs portal coupling, such as $\kappa=0.148$ (BP-c4) are mainly dominated by the annihilation through $s+cross$-channels. However, the relic density for the point BP-c3 is coming due to the combined contributions of $s+cross$-channels and $t+u$-channels.
\begin{table*}[h!]
	\centering
	\begin{tabular}{|p{1.6cm}|p{1.2cm}|p{1.1cm}|p{1.2cm}|p{1.2cm}|c|p{4.7cm}|}
		\hline
		\hline
		Channel & $M_{DM}$ (GeV) & ~~$\kappa~~~$& $M_{E_1^\pm}$ (GeV) &~~$Y_{f}$&$\Omega_{DM}h^2$&~~~~~~~~~~Percentage \\
		\hline
		&&&&&&$\sigma(SS\rightarrow \nu \nu)\quad 72\%$  \\
		~~BP-d1&325&0.05&1000& 0.225 &0.1173& $\sigma(SS\rightarrow W^\pm W^\mp) \quad 12\%$ \\
		&&&&&& $\sigma(SS\rightarrow HH)\quad7\%$\\
		&&&&&& $\sigma(SS\rightarrow ZZ)\quad6\%$\\
		&&&&&&$\sigma(SS\rightarrow t\bar{t})\quad 4\%$ \\
		\hline
		&&&&&&$\sigma(SS\rightarrow \nu \nu)\quad 88\%$ \\
		~~BP-d2&500&0.05&1000& 0.250 &0.1219&$\sigma(SS\rightarrow W^\pm W^\mp) \quad5\%$\\
		&&&&&& $\sigma(SS\rightarrow ZZ)\quad3\%$\\
		&&&&&& $\sigma(SS\rightarrow HH)\quad3\%$\\
		\hline
		&&&&&&$\sigma(SS\rightarrow \nu \nu)\quad 96\%$ \\
		~~BP-d3&675&0.05&1000& 0.280 &0.1169& $\sigma(SS\rightarrow W^\pm W^\mp) \quad3\%$\\
		&&&&&& $\sigma(SS\rightarrow ZZ)\quad1\%$\\
		&&&&&& $\sigma(SS\rightarrow HH)\quad1\%$\\
		\hline
		\hline
	\end{tabular}
	\caption{The benchmark points allowed by all the theoretical and experimental constraints. $\sigma(SS\rightarrow \nu \nu)$ is mainly dominated by the $t+u$-channel annihilation processes whereas $\sigma(SS\rightarrow YY),\,Y=W,Z,H,t$ dominated by the  $s+cross$-channel annihilation processes. We consider $Y_{f1}=Y_{f3}=Y_f$ to avoid flavour violating decay processes.}
	\label{tabDM:3}
\end{table*}
We also scan in the other three dimensional parameter space. The dark matter mass $M_{DM}$ is varied from 5 GeV to 540 GeV and $\kappa$ from $-0.35$ to $0.35$ with a step size $0.002$ and $Y_f$ from $-0.35$ to $0.35$ GeV with a step size $0.005$ GeV with fixed $M_{E_1^\pm}=500$ GeV. It is noted that the co-annihilation effect are completely absent here as $\Delta M^{\pm,0}> 0.1 M_{DM}$. We display the allowed parameters $\kappa-M_{DM}$ plane in Fig.~\ref{fig:DarkAll1}.
One can see, in the presence of DM annihilation via $t,u$-channel as most of the region is giving the correct DM density, which is also allowed by other experimental constraints. For $\kappa\neq 0$, the $s$-channel annihilation dominates near Higgs resonance region $\sim \frac{M_H}{2}$. This region gives overabundance of dark matter density in our study. For a small $\kappa \sim 0$, the $t+u$-channels helps to get the correct relic density at $3\sigma$ C.L.
We show the $\kappa-Y_f$ plane in Fig.~\ref{fig:DarkAll2} for the same data points as in Fig.~\ref{fig:DarkAll1}. We get two circular ring-type structures here. The empty region violates one of the constraints, such as the relic density of the dark matter, direct detection, and Higgs decay width for the DM mass $<\frac{M_H}{2}$. 
Nonetheless, in presence of the co-annihilation processes with/or a different choice of the $M_{E_1^\pm}$, the gaps between these two circular rings could be filled. We also display similar plots in $\kappa-M_{DM}$ and  $\kappa-Y_f$ planes in Figs.~\ref{fig:DarkAll3} and~\ref{fig:DarkAll4} for the  $M_{E_1^\pm}=1000$ GeV, where we change the variation for DM mass $M_{DM}$ from 5 GeV to 1000 GeV. We get a similar type of plot with a large region of the parameter spaces allowed by all the experimental and theoretical constraints. Few BMPs and their corresponding contributions are presented in Tab.~\ref{tabDM:3}. $\sigma(SS\rightarrow \nu \nu)$ is mainly dominated by the $t+u$-channel annihilation processes whereas $\sigma(SS\rightarrow YY),\,Y=W,Z,H,t$ dominated by the  $s+cross$-channel annihilation processes.   
\subsection{Neutrino mass and mixing}
In this minimal model, with the choice of parameter space we discuss some numerical insights to neutrino phenomenology.
Using equation \eqref{eq:n1}, with the masses for subsequent fields $M_{DM}=110$ GeV $M_N=800$ GeV and  choice of Yukawa parameters $|Y_{f1}|=0.8 |,~ Y_{f2}|=10^{-4},|Y_{f3}|=0.282$, we get the sum of the neutrino masses of the order of sub-eV range ($\sim0.03$ eV) for Higgs portal coupling $\kappa<10^{-6}$. This smallness of neutrino mass does satisfy current upper bound on sum of the active neutrino masses \cite{Giusarma:2016phn, Vagnozzi:2017ovm}, and the tiny $\kappa$ is also directly associated with dark matter relic density via the $t$-channel process.
We are able to generate mixing angles $\theta_{12}=32.7^{\circ}$, $\theta_{13}=8.4^{\circ}$, $\theta_{23}=44.71^{\circ}$ and mass differences $\Delta m_{21}^2=7.44\times10^{-5}$ eV$^2$ and $\Delta m_{31}^2=4.9\times10^{-4}$ eV$^2$ with phases $\alpha=\delta=45^{\circ}$. Although $\Delta m_{21}^2$ is within the present 3$\sigma$ bound yet, $\Delta m_{31}^2$ is deviate from the actual range for this choice of the parameters. For the other choice of the parameters (mainly $Y_{f2}$), 
we can get $\Delta m_{31}^2$ is within the present 3$\sigma$ bound yet, $\Delta m_{21}^2$ is deviate from the actual range for this choice of the parameters. We have also opted for Casas-Ibarra parametrization \cite{Casas:2001sr} extended to radiative model \cite{Toma:2013zsa} to solve this inconsistency, however, in that case also exactly similar kind of problem arises. As a conclusive remark we obtained from this study is that, within the scotogenic model, it is unable to explain all the neutrino oscillation parameters with just a single generation of fermion doublet. One must introduce an additional field that interact with the lepton doublet to successfully explain the neutrino mass. 
We are adding an extra fermion doublet $F$ as {\it ad hoc} basis in the model with mass $M_F=2500$ GeV to test the inconsistency. The interaction Lagrangian of Eq. \eqref{lint} will be slightly modified as  $\sum\limits_{{\scalebox{0.5}{i={1,2,3}}, ~ \scalebox{0.5}{j=1,2}}} Y_{fij} \bar{L_i} F_{Dj} S $. The Yukawa couplings are set as $|Y_{f11}|=0.1, |Y_{f12}|=5\times10^{-4}, |Y_{f13}|=5\times10^{-3}, |Y_{f21}|=4\times 10^{-2}, |Y_{f22}|=4.2\times 10^{-3}$ and $|Y_{f23}|=4.9\times 10^{-2}$ to observed exact 3$\sigma$ bounds on the light neutrino parameters. These set of couplings give rise to  $\Delta m_{21}^2=7.08\times10^{-5}$ eV$^2$ and $\Delta m_{31}^2=2.5\times10^{-3}$ eV$^2$ with phases $\alpha=28.6479^{\circ}$ and $\delta=42.9718^{\circ}$, which satisfies the current bound on the parameters space \cite{Esteban:2020cvm}. Even though the inclusion of this {\it ad hoc} particle with low mass could explain the neutrino parameters completely, anyhow it could also affect dark matter parameter space in this model itself. However we check that the choice of $M_F=2500$ GeV and with a small mixing parameters from $\sum\limits_{\scalebox{0.5}{j=1,2}} Y_{N_j}\overline{F}_D\phi E_S$ have a very tiny effect on the relic density calculation. It will also have negligibly small affect on the collider search results for first generation of heavy fermion. To get the collider signature of this heavy particle we need a very large luminosity $>1~{\rm ab^{-1}}$ (ab=attobarn) at LHC with energy $\sqrt{s}=14$ TeV.
\subsection{Collider Searches}
We perform a search for the lightest charged fermion $E_1^\pm$ in the context of 14 TeV LHC experiments with integrated luminosity of 3000 fb$^{-1}$ for event's process $ pp \rightarrow E_1^\pm E_1^\mp$,  where a SM leptons $l$ is produced through decays of the charged fermion as $E_1^\pm\rightarrow l^\pm S$. Hence, in the final state, events have two
same flavours opposite sign (SFOS) leptons, including significant missing transverse energy coming from the LSP $S$. 
Here, processes like $pp \rightarrow WW~ (W \rightarrow l \nu)$, $pp\rightarrow ZW~(Z \rightarrow ll, W\rightarrow l \nu )$ and $pp\rightarrow ZZ ~(Z \rightarrow ll, Z\rightarrow \nu \overline{\nu})$ can add to the SM background if additional charged leptons get misidentified or remain unreconstructed. Also other reducible backgrounds like $pp\rightarrow t\overline{t}, t \rightarrow W b, W \rightarrow l \nu$ may also produce two leptons and jets in the final state.
Similarly the Drell–Yan process $pp \rightarrow Z^*, \gamma^* \rightarrow ll ~jets$ ($jets$ misidentified in our case) can also contribute to the SM background.
Multi-jets final state events would hugely affected by this background due to the large production cross-section. The additional cuts on number of jets reduce these backgrounds to be less than one. This channel $ pp \rightarrow E_1^\pm E_1^\mp, E_1^\pm\rightarrow l^\pm S \rightarrow ll + \slashed{E}_T$ is analyzed by performing a detailed cut based collider analysis. We will show the projected exclusion/discovery reach of direct heavy charged fermion and dark matter as transverse missing energy searches in this channels by performing a detailed cut based collider analysis.

We use {\tt FeynRules}~\cite{Alloul:2013bka} to get the input codes for  {\tt MadGraph-2.6.5}~\cite{Alwall:2014hca}. Using the particle spectrum into the {\tt MadGraph-2.6.5}, we calculate the production cross-section of the heavy charged fermions.
We also verified the results using {\tt SARAH-4.8.6}~\cite{Staub:2012pb,Staub:2015kfa} including {\tt SPheno-4.0.3}~\cite{Porod:2011nf} mass spectrum into the {\tt MadGraph-2.6.5}. We use {\tt MadGraph-2.6.5} to generate the
signal as well as background events and {\tt  PYTHIA-8.2}~\cite{Sjostrand:2014zea} for showering and hadronization. All generated signal and background events are processed through a fast simulation package {\tt Delphes-3.4.1}~\cite{deFavereau:2013fsa} and we choose ALTAS configuration card for the analysis.

\begin{figure}
\begin{center}
\includegraphics[scale=0.63]{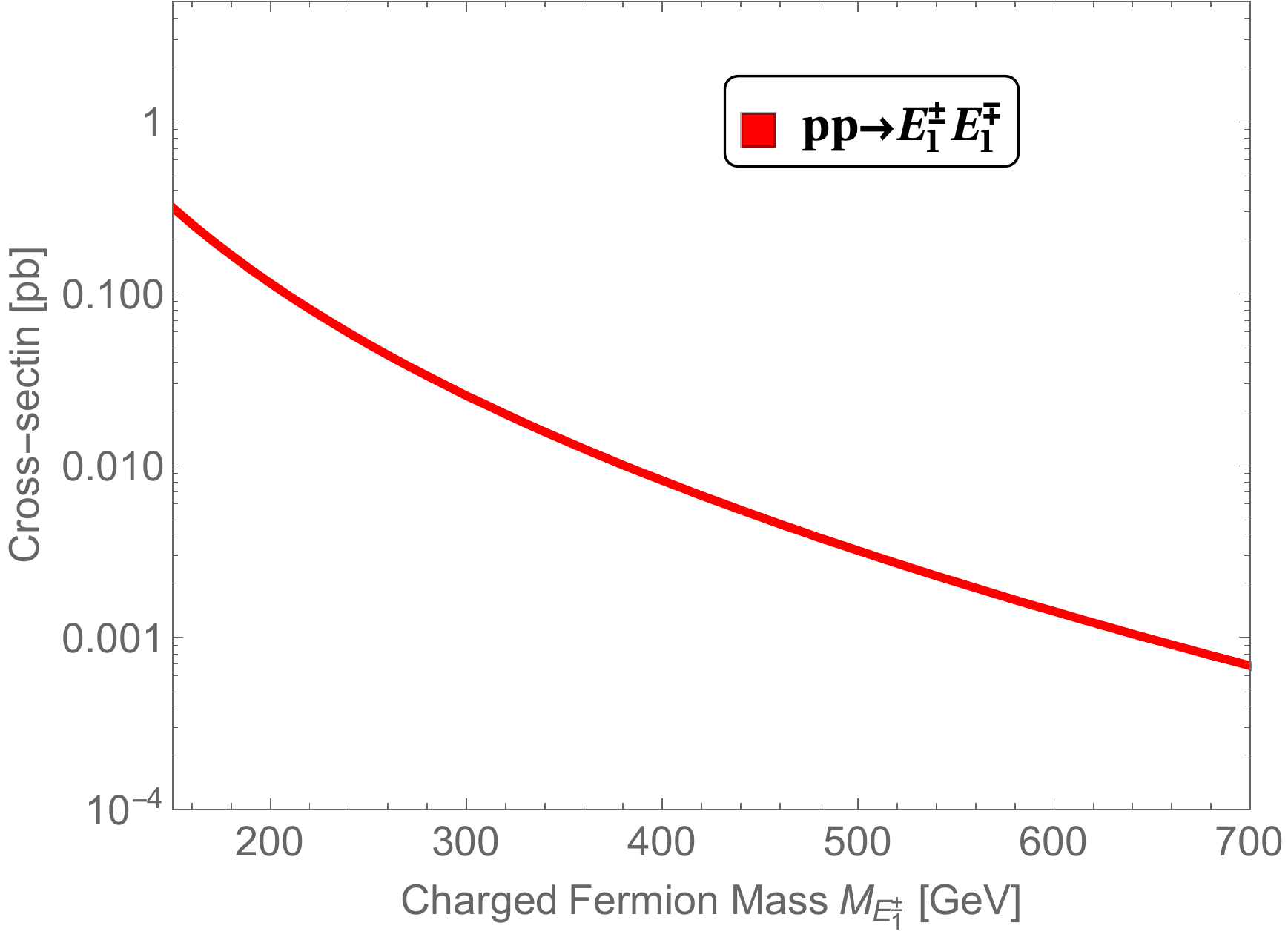}
\end{center}
\caption{Production cross-section for the heavy charged fermions $ pp \rightarrow E_1^\pm E_1^\mp$ at the 14 TeV LHC experiment.}
\label{fig:crossMe1}
\end{figure}

The events are selected with two same flavours opposite sign (SFOS) isolated electron. Total number of muon is very less in the final state events as $Y_{f2}\sim 0$. Anyways, we consider both the electron and muon in the final state with transverse momentum $p_T$ larger than 30 GeV. The charged lepton isolation requires that there is no other charged particle with $p_T > 0.5$ GeV/c within a cone of
$\Delta R = \sqrt{\Delta\Phi^2 +\Delta \eta^2} < 0.5$ centered on the cell-associated to the charged lepton. Besides, the ratio of the scalar sum of the transverse momenta of all tracks to $p_T$ of the lepton (chosen for isolation) is less than $0.12$ ($0.25$) for the electron (muon). Here $p_T$, $\Phi$ and $\eta$ are the transverse momentum, polar angle and pseudo-rapidity of charged leptons respectively.
The charged lepton candidates  are required to be within a pseudorapidity range of $|\eta| < 2.5$. Number of light and $b$-jets in the final state are taken to be zero.

A variety of kinematic variables have been used to design the optimized signal regions. Out of them, the invariant mass $M_{ll}$ and  transverse missing energy $\slashed{E}_T$ can be a
useful probe to search for the charged fermion $E_1^\pm$ of this model.
First and foremost, the invariant mass of the two final state $l$, $M_{ll} \equiv M_{l_{1}l_{2}}$ ($l_{1}$ and $l_{2}$ represents the $p_{T}$ ordered leading and sub-leading $l$ in the final state) is used to discriminate the background.
For the signal process, the $l$ pair is produced from the decay of the $E_1^\pm$ and thereby no peaks around $\sim 110-115~{\rm GeV}$. On the other hand, the $M_{ll}$ distribution for the most dominant $ZZ/WZ$ background has a peak roughly around $\sim 110-115~{\rm GeV}$ and the distribution smoothly falling after $115$ GeV since the two $l$ are produced from the decay of $Z$ boson.
The transverse missing energy $\slashed{E}_T$ can also be a
useful probe to search for the charged fermion $E_1^\pm$ of this model where the only contribution to $\slashed{E}_T$ comes from the $\nu$ produced by the leptonically decaying $W$ for the background and the dark matter itself behave as $\slashed{E}_T$ for the signal process.
\begin{table}[htpb!]
\begin{center}\scalebox{1.0}{
\begin{tabular}{|c|c|c|c|c|c|c|c|c|c|c|c|c|c|}
\hline
\hline
Signal Region &\multicolumn{2}{c|}{ Various Cuts   }\\
\cline{2-3}
& ~~~~$M_{ll}$ [GeV]~~~~&  $~~~~\slashed{E}_T$ [GeV]~~~~\\
\cline{1-3}
\rule{0pt}{1ex}
SR-1 & 100.0 & 100.0 \\
\rule{0pt}{1ex}
SR-2  & 110.0 & 140.0  \\
\rule{0pt}{1ex}
SR-3  & 120.0 & 200.0 \\
\hline
\hline
\end{tabular}}
\end{center}
\caption{The three optimized signal regions (SR-1, SR-2 and SR-3).}
\label{table:sr}
\end{table}

Signal events have been generated for various combinations of $M_{E_1^\pm}$ and $M_S$. $M_{E_1^\pm}$ has been varied from $150~{\rm GeV}$ to $1600~{\rm GeV}$ with a step size of $30~{\rm GeV}$, while $M_S$ has been varied between $10~{\rm GeV}$ to $M_{E_1^\pm}$ with a step size of $30~{\rm GeV}$. It is to be noted that to avoid the bounds from the Electroweak Precision Parameters $S,T$ and $U$ we took large $\cos\beta=0.995$ and $M_{E_2^\pm}=1500$ GeV and fixed  $Y_f=0.165$. 

\begin{table}[htpb!]
\begin{center}\scalebox{1.0}{
\begin{tabular}{|c|c|c|c|c|c|c|c|c|c|c|c|c|c|}
\hline
\hline
BMPs & Cross-Sections [fb] & Backgrounds & Cross-Sections [fb]\\
\cline{1-4}
\rule{0pt}{1ex}
BMP-1 & 303.1  & $ pp \rightarrow WW $&  28.2102\\
\rule{0pt}{1ex}
BMP-2  & 4.806 & $ pp \rightarrow WZ$ & 12.5581\\
\rule{0pt}{1ex}
BMP-3  & 2.91 & $ pp \rightarrow ZZ$ &30.0432\\
\hline
\hline
\end{tabular}}
\end{center}
\caption{The cross-section for the process $ pp \rightarrow E_1^\pm E_1^\mp, E_1^\pm\rightarrow l^\pm S$ for three benchmark points (BMP-1, BMP-2 and BMP-3).}
\label{tab:cs}
\end{table}


Three different signal regions are chosen, SR-1, SR-2 and SR-3 (see table~\ref{table:sr}), aimed at maximizing the significance of signal events with small, intermediate and large mass difference, respectively~\cite{Barman:2020azo}. 
The selection cuts for SR-1, SR-2 and SR-3 have been chosen by performing a cut based analysis for the three representative benchmark points: BMP-1: $M_{E_1^\pm}= 150~{\rm GeV}$, $M_S = 50~{\rm GeV}$, BMP-2: $M_{E_1^\pm}= 450~{\rm GeV}$, $M_S = 300~{\rm GeV}$ and BMP-3: $M_{E_1^\pm}= 500~{\rm GeV}$, $M_S = 30~{\rm GeV}$, respectively. 
\begin{table}[htpb!]
\begin{center}\scalebox{1.0}{
\begin{tabular}{|c|c|c|c|c|c|c|c|c|c|c|c|c|c|}
\hline
\hline
Signal&Total &\multicolumn{6}{c|}{ Benchmark points: ($M_{E_1^\pm}$, $M_S$) in GeV   }\\
\cline{3-8}
Region&number of& \multicolumn{2}{c|}{  BMP-1~(150, 50) }&\multicolumn{2}{c|}{  BMP-2~(450, 300) }&\multicolumn{2}{c|}{ BMP-3~(500, 30) }\\
\cline{3-8}
&Backgrounds& ~~$\#$ events~~& ~~Significance&$\#$ events~~& ~~Significance&$\#$ events~~& ~~Significance\\
\cline{1-8}
\rule{0pt}{1ex}
SR-1 & 5260.9856& 52768.92 & 178.414 & 2296.05 & 23.42 & 1883.05 & 19.6\\
\rule{0pt}{1ex}
SR-2 & 2361.8417 & 10320.5& 91.64 & 1613.35 & 25.5887 & 1718.07& 26.9\\
\rule{0pt}{1ex}
SR-3 &668.5 & 1909.42 & 37.6 & 705.755 &19.038 & 1464.02 & 31.6978\\
\hline
\hline
\end{tabular}}
\end{center}
\caption{The signal significance for the three benchmark signal points ( BMP-1, BMP-2 and BMP-3) corresponding to the three optimized signal regions (SR-1, SR-2 and SR-3) are shown. In addition, the total background yield and the total signal yield is also given.}
\label{table:signal_significance}
\end{table}

We show the cross-section of the process $ pp \rightarrow E_1^\pm E_1^\mp$ for different mass in the Fig.~\ref{fig:crossMe1}.
The values of $\sigma_{pp \to E_1^\pm E_1^\mp, E_1^\pm\rightarrow l^\pm S}$ for BMP-1, BMP-2 and BMP-3 have been listed in Table~\ref{tab:cs}.
The signal yield ($S$) has been computed as follows:
\begin{eqnarray}
S = \sigma_{pp \to E_1^\pm E_1^\mp, E_1^\pm\rightarrow l^\pm S} \times  \mathcal{L} \times \varepsilon_{Eff}.
\label{Eqn:signal_yield} 
\end{eqnarray}
where, $\mathcal{L}$ is the integrated future LHC luminosity ($\mathcal{L}=3000~ {\rm fb^{-1}}$) and $ \varepsilon_{Eff}$ represents the efficiency of the signal region. $ \varepsilon_{Eff}$ is the ratio of the number of signal events which pass through a certain signal region ($\rm N_{initial}$) to the total number of generated signal events ($\rm N_{final}$); $ \varepsilon_{Eff}$ = ${\rm  N_{final}/ N_{initial}}$.

 \begin{figure}[h!]
	\begin{center}{
			\includegraphics[scale=0.55]{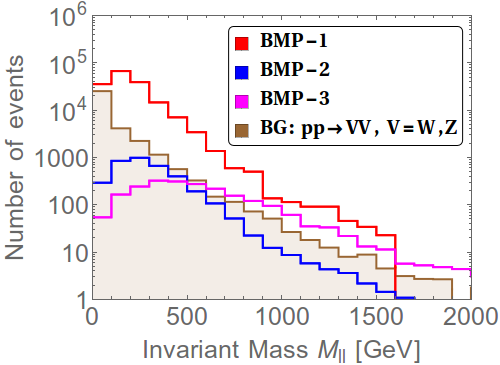}}
		\caption{ \rm The invariant mass distribution of the two same flavour opposite sign (SFOS) leptons for the signal $ pp \rightarrow E_1^\pm E_1^\mp, E_1^\pm\rightarrow l^\pm S \rightarrow ll + \slashed{E}_T$ and $ pp \rightarrow VV, V=W,Z$ backgrounds. The distributions for BMP-1, BMP-2 and BMP-3 have been illustrated as red, blue and magenta solid colors while the $pp \rightarrow VV, V=W,Z$ background has been shown in brown color.}
		\label{fig:coll1}
	\end{center}
\end{figure}
 \begin{figure}[h!]
	\begin{center}{
			\includegraphics[scale=0.55]{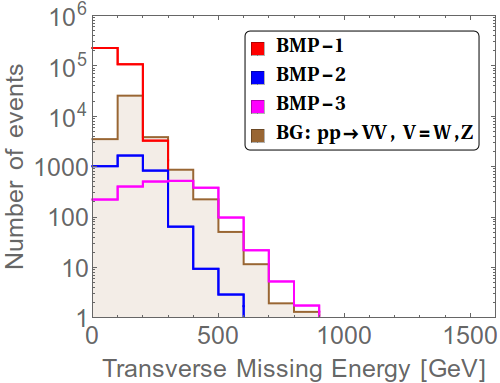}}
		\caption{ \rm The transverse mass energy distribution for the signal $ pp \rightarrow E_1^\pm E_1^\mp, E_1^\pm\rightarrow l^\pm S \rightarrow $ $ ll + \slashed{E}_T$ and $ pp \rightarrow VV, V=W,Z$ backgrounds. The distributions for BMP-1, BMP-2 and BMP-3 have been illustrated as red, blue and magenta solid colors while the $pp \rightarrow VV, V=W,Z$ background has been shown in brown color.}
		\label{fig:coll2}
	\end{center}
\end{figure}

The $M_{ll}$ distribution for the signal benchmark points (BMP-1, BMP-2 and BMP-3) and the total background $pp \rightarrow VV, V=W,Z$ has been shown in Figs.~\ref{fig:coll1}.
The distributions for BMP-1, BMP-2 and BMP-3 in Figs.~\ref{fig:coll1} have been illustrated as red, blue and magenta solid colors while the $pp \rightarrow VV, V=W,Z$ background has been shown in brown color.  
 We show the transverse missing energy $\slashed{E}_T$ distributions in Fig.~\ref{fig:coll2} for the same benchmark points.

 \begin{figure}[h!]
	\begin{center}{
			\includegraphics[scale=0.5]{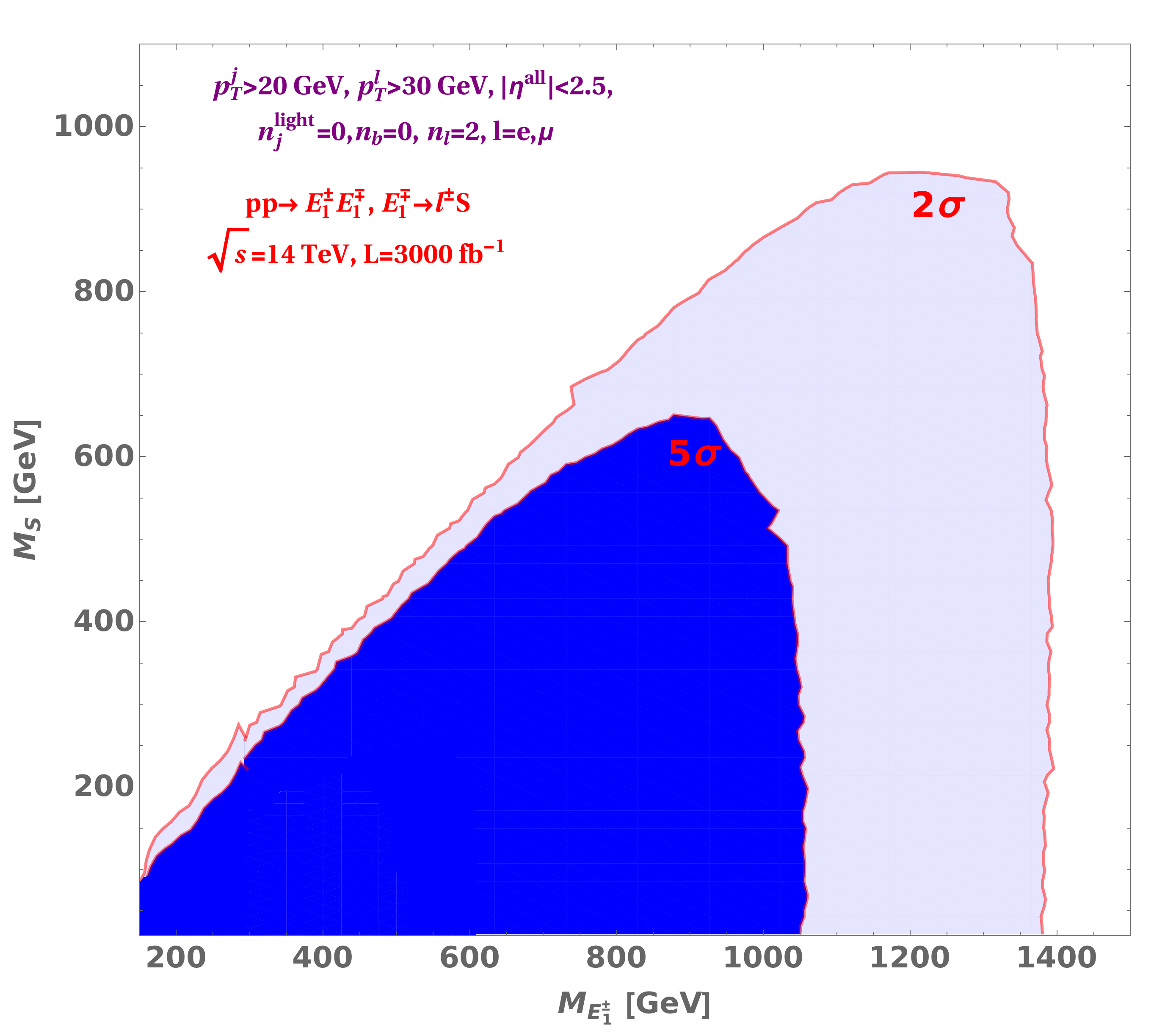}}
		\caption{ \rm The exclusion plot for the signal $ pp \rightarrow E_1^\pm E_1^\mp, E_1^\pm\rightarrow l^\pm S \rightarrow $ $ ll + \slashed{E}_T$ and $ pp \rightarrow VV, V=W,Z$ backgrounds at $\sqrt{s}=14$ TeV with integrated luminosity $L=3000~{\rm fb^{-1}}$.}
		\label{fig:cont}
	\end{center}
\end{figure}

The optimized selection cuts corresponding to SR-1, SR-2 and SR-3 are given in table~\ref{table:sr}. The signal yields for BMP-1, BMP-2 and BMP-3, along with the corresponding background yields obtained after application of selection cuts listed in SR-1, SR-2 and SR-3, respectively, have also been shown in Table~\ref{table:signal_significance}. It should be noted that the signal significances have been obtained without assuming any systematic uncertainty. The signal significance is computed as $S/\sqrt{S+B}$, where $S$ and $B$ are the signal and background yields. SR-1 results in a signal significance of $178.414$ for BMP-1, while SR-2 and SR-3 has a signal significance of $25.5887$ and $31.6978$ for BMP-2 and BMP-3, respectively.

We also derive the projected exclusion limits in the $M_{E_1^\pm}-M_S$ plane from direct heavy charged-fermion searches at 14 TeV LHC experiments with an integrated luminosity of 3000 ${\rm fb^{-1}}$ in the $2l+\slashed{E}_T$ search channel. The value of signal significance is computed for the three optimized signal regions (SR-1, SR-2 and SR-3) and the maximum among them is considered in deriving the projection regions. The projected exclusion and discovery region corresponds to the sector with signal significance $> 2\sigma$ and $> 5\sigma$, respectively. They have been represented in light blue and dark blue colors, respectively, in Fig.~\ref{fig:cont}.
 It can be observed from Fig.~\ref{fig:cont} that in this model, direct heavy charged-fermion searches at 14 TeV LHC experiments with an integrated luminosity of 3000 ${\rm fb^{-1}}$ in the $2l+\slashed{E}_T$ search channel has a potential exclusion (discovery) reach up to $\sim 1380~{\rm GeV}$ ($\sim 1050~{\rm GeV}$) for the dark matter mass $M_S <10~{\rm GeV}$.
 It is to be noted that the stransverse $m_{T2}$~\cite{Lester:1999tx, Cho:2007qv} cut in this analysis is not included, and we also agree if we include it may exclude larger region of the parameter space. The ATLAS~\cite{Aad:2019vnb} collaboration done such analysis including $m_{T2}$ cut from direct slepton searches in the $ ll + \slashed{E}_T$ final state at $\sqrt{s}=13$ TeV with integrated luminosity $\mathcal{L}=3000~{\rm fb^{-1}}$ within a simplified R-parity conserving Supersymmetry framework.
It is also true that one can include further kinematic variables (such as different $p_T,\Delta \eta_{ll}, \Delta \Phi_{ll}$ etc. in different region) and/or more signal regions to maximize the exclusion contour on the same plane. Further analysis by including more number of cuts may lead to a better and improved results. However in this simplified model, the minimum sets of cut $M_{ll}$ ans $\slashed{E}_T$ with three signal region are enough to probe this model at future collider analysis.

 \begin{figure}[h!]
	\begin{center}{
	        \includegraphics[scale=0.5]{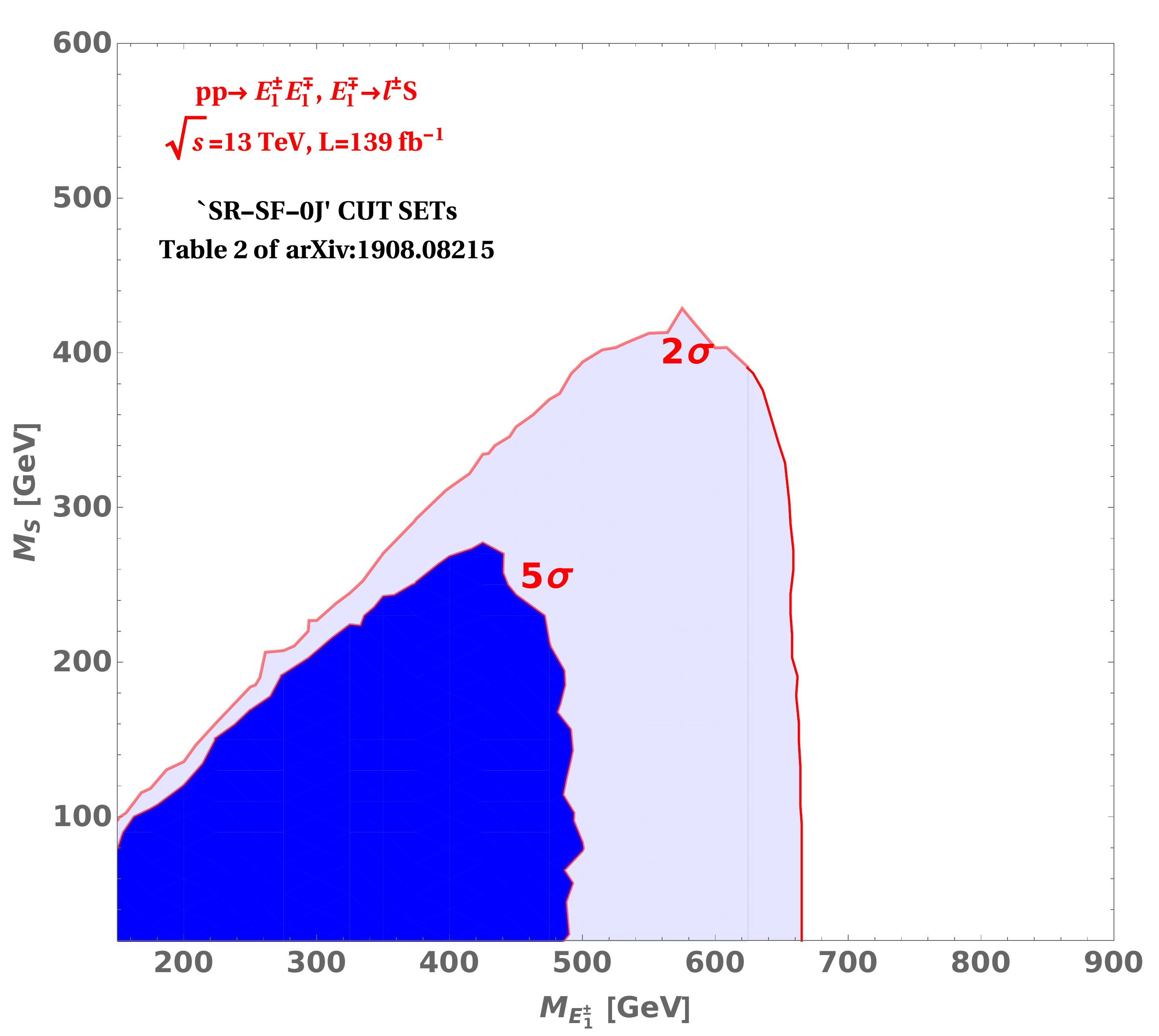}}
		\caption{ \rm The exclusion plot for the signal $ pp \rightarrow E_1^\pm E_1^\mp, E_1^\pm\rightarrow l^\pm S \rightarrow $ $ ll + \slashed{E}_T$ and $ pp \rightarrow VV, V=W,Z$ backgrounds at $\sqrt{s}=13$ TeV with integrated luminosity of 139 fb$^{-1}$.}
		\label{fig:cont3000s13}
	\end{center}
\end{figure}

It would also be good to see, how much this model is already probed by current
data, following the outline in the ATLAS search at $\sqrt{s}=13$ TeV with integrated luminosity 139 fb$^{-1}$~\cite{Aad:2019vnb}. Including the invariant mass $M_{ll}$ and transverse missing energy $\slashed{E}_T$, they used the stransverse $m_{T2}$~\cite{Lester:1999tx, Cho:2007qv} and other cuts (see Table 2 of the Ref.~\cite{Aad:2019vnb}). They choose various final state.  To compare with our analysis, we stick with ``SR-SF-0J" (same flavor opposite sign leptons without $jets$) final state. We generate the events for different combinations of $M_{E_1^\pm}$ and $M_S$. $M_{E_1^\pm}$ has been varied from $150~{\rm GeV}$ to $800~{\rm GeV}$ with a step size of $40~{\rm GeV}$, while $M_S$ has been varied between $10~{\rm GeV}$ to $M_{E_1^\pm}$ with a step size of $30~{\rm GeV}$.
For the same signal region as in Table 2 of the Ref.~\cite{Aad:2019vnb}, one can also get the similar  exclusion contour plot (shown in Fig.~\ref{fig:cont3000s13}) for the direct heavy charged-fermion searches at 13 TeV LHC experiments with an integrated luminosity of 139 fb$^{-1}$. We find the exclusion (discovery) reach up to $\sim$ 700 GeV
 ($\sim$ 500 GeV) for the dark matter mass $M_S < 10$ GeV.

\section{Conclusion}\label{conc}
In this work, we study the possibility of singlet scalar dark matter and neutrino mass in the minimal scotogenic model. The structure of the model projected here uses a minimum number of the field content. On the top of the SM field content, this model contains vector-like one neutral and two charged fermions along with a singlet scalar field. With the minimal choice of field content, we are unable to afford all the neutrino oscillation parameters at the right ballpark. Hence, one extra fermion doublet is added in the model as {\it ad hoc} basis to complete the neutrino framework. In the presence of other particles, one can get the correct relic density via co-annihilation, or one may have the interaction term such that the dark matter can annihilate into SM particles through additional cross-channel, $t$- and $u$-channels. The constructive or destructive interference among these channels helps to modify the effective annihilation cross-section and give the right relic density of the dark matter in our model.

The vector-like fermions have an interaction term with the Higgs scalar fields for which give rise to a mass difference between the degenerate neutral and charged fermions of the doublet at tree-level. The interaction term with singlet scalar helps to generate the neutrino mass and mixing angles via a 1-loop level through the radiative seesaw mechanism. Both of these interactions terms also help to get the exact relic density of the Universe for large ranges $0.1-100$ TeV ($M_{DM}\gtrsim 100$ TeV violates the unitary bounds~\cite{Griest:1989wd, Smirnov:2019ngs}) of the dark matter mass. The Higgs portal coupling $\mathcal{O}(10^{-5})$ along with these Yukawa couplings $\mathcal{O}(10^{-1})$ can explain the neutrino mass and mixing angles where the relic density is achieved via the $t$- and $u$-channel annihilation or other co-annihilation processes.
These new Yukawa couplings also play the lead role in explaining the discrepancy of the muon anomalous magnetic moment.

 We also performed collider analysis to search the lightest charged fermion $E_1^\pm$ in the context of 14 TeV LHC experiments with integrated luminosity of 3000 fb$^{-1}$ for process $ pp \rightarrow E_1^\pm E_1^\mp$  where, a SM leptons $l$ is produced through decays of the charged fermion as $E_1^\pm\rightarrow l^\pm S$. 
We have only analyzed the familiar $2l+\slashed{E}_T$ final states to get the signature at the future collider. The leptonic final states produce relatively clean signals which are easy to identify in a hadron-rich environment like the LHC experiment. We choose benchmark points that ensure the relic density and neutrino parameters. We further optimized the selection cuts to enhance the $2l+\slashed{E}_T$ signal significance over the SM backgrounds.   Our collider study showed that the dilepton final state gives promising results for the discovery of the heavy charged particle at 14 TeV LHC experiments with an integrated luminosity of 3000 ${\rm fb^{-1}}$, which may be an indication of the dark matter at the collider.
We have also shown the projected exclusion and discovery limits in the $M_{E_1^\pm}-M_S$ plane from direct heavy charged-fermion searches at 14 TeV LHC experiments with an integrated luminosity of 3000 ${\rm fb^{-1}}$ in the $2l+\slashed{E}_T$ search channel. The projected exclusion and discovery region corresponds to the sector with signal significance $> 2\sigma$ and $> 5\sigma$, respectively.
 It can be observed that in this model, direct heavy charged-fermion searches at 14 TeV LHC experiments with an integrated luminosity of 3000 ${\rm fb^{-1}}$ in the $2l+\slashed{E}_T$ search channel has a potential exclusion (discovery) reach up to $\sim 1380~{\rm GeV}$ ($\sim 1050~{\rm GeV}$) for the dark matter mass $M_S <10~{\rm GeV}$.

One can also put bound on the Yukawa coupling as larger Yukawa coupling may violate the stability of the scalar potential any of the direction the scalar fields at any scale (at least up to the Planck scalar $1.22 \times 10^{19}$ GeV). In this model, we work with such a choice of the Yukawa couplings and $\kappa$ (especially $\lambda_S$) so that there is no new minima arise along any of the scalar field directions. In the future, we will elaborate on the details stability and/or metastability analysis for various regions of the parameter space, which could also explain all the neutrino masses and mixing angles, exact relic density and baryon-asymmetry of the Universe altogether.

In the concluding remark: if nature selects a single component WIMP dark matter candidate, which interacts with the nucleus feebly through $s$-channel, helps to get the neutrino mass of order $\mathcal{O}(0.1)$ eV. On the assumption that the relic density can achieve via  $t$-channel annihilation processes and we may have to think of a new way to detect dark matter in the direct-detection experiments. In that case, collider searches with high luminosity are better options to detect dark matter.
 \section{Acknowledgement}
The authors would like to acknowledge Narendra Sahu from IIT Hydrabad for fruitful discussion. The research work of P.D. and  M.K.D. is supported by the Department of Science and Technology, Government of India under the project grant EMR/2017/001436. NK would like to thank Biplob Bhattacharjee for the help on collider analysis and to Dilip Kumar Ghosh for his support at IACS.
\bibliographystyle{apsrev4-1}
\bibliographystyle{utphys}
\bibliography{tevportalnew}

\providecommand{\href}[2]{#2}\begingroup\raggedright\begin{thebibliography}{10}

\bibitem{Sirunyan:2017khh}
{\bfseries CMS} Collaboration, A.~M. Sirunyan {\em et~al.}, ``{Observation of
  the Higgs boson decay to a pair of $\tau$ leptons with the CMS detector},''
  \href{http://dx.doi.org/10.1016/j.physletb.2018.02.004}{{\em Phys. Lett.}
  {\bfseries B779} (2018) 283--316},
\href{http://arxiv.org/abs/1708.00373}{{\ttfamily arXiv:1708.00373 [hep-ex]}}.

\bibitem{Sirunyan:2018koj}
{\bfseries CMS} Collaboration, A.~M. Sirunyan {\em et~al.}, ``{Combined
  measurements of Higgs boson couplings in proton–proton collisions at
  $\sqrt{s}=13\,\text {Te}\text {V} $},''
  \href{http://dx.doi.org/10.1140/epjc/s10052-019-6909-y}{{\em Eur. Phys. J.}
  {\bfseries C79} no.~5, (2019) 421},
\href{http://arxiv.org/abs/1809.10733}{{\ttfamily arXiv:1809.10733 [hep-ex]}}.

\bibitem{Tanabashi:2018oca}
{\bfseries Particle Data Group} Collaboration, M.~Tanabashi {\em et~al.},
  ``{Review of Particle Physics},''
\href{http://dx.doi.org/10.1103/PhysRevD.98.030001}{{\em Phys. Rev.} {\bfseries
  D98} no.~3, (2018) 030001}.

\bibitem{ArkaniHamed:2008qn}
N.~Arkani-Hamed, D.~P. Finkbeiner, T.~R. Slatyer, and N.~Weiner, ``{A Theory of
  Dark Matter},'' \href{http://dx.doi.org/10.1103/PhysRevD.79.015014}{{\em
  Phys. Rev.} {\bfseries D79} (2009) 015014},
\href{http://arxiv.org/abs/0810.0713}{{\ttfamily arXiv:0810.0713 [hep-ph]}}.

\bibitem{Dasgupta:2014hha}
A.~Dasgupta and D.~Borah, ``{Scalar Dark Matter with Type II Seesaw},''
  \href{http://dx.doi.org/10.1016/j.nuclphysb.2014.10.028}{{\em Nucl. Phys.}
  {\bfseries B889} (2014) 637--649},
\href{http://arxiv.org/abs/1404.5261}{{\ttfamily arXiv:1404.5261 [hep-ph]}}.

\bibitem{Babu:2009fd}
K.~S. Babu, \href{http://dx.doi.org/10.1142/9789812838360_0002}{``{TASI
  Lectures on Flavor Physics},''} in {\em {Proceedings of Theoretical Advanced
  Study Institute in Elementary Particle Physics on The dawn of the LHC era
  (TASI 2008): Boulder, USA, June 2-27, 2008}}, pp.~49--123.
\newblock 2010.
\newblock
\href{http://arxiv.org/abs/0910.2948}{{\ttfamily arXiv:0910.2948 [hep-ph]}}.
\newblock

\bibitem{Burgess:2000yq}
C.~P. Burgess, M.~Pospelov, and T.~ter Veldhuis, ``{The Minimal model of
  nonbaryonic dark matter: A Singlet scalar},''
  \href{http://dx.doi.org/10.1016/S0550-3213(01)00513-2}{{\em Nucl. Phys.}
  {\bfseries B619} (2001) 709--728},
\href{http://arxiv.org/abs/hep-ph/0011335}{{\ttfamily arXiv:hep-ph/0011335
  [hep-ph]}}.

\bibitem{Deshpande:1977rw}
N.~G. Deshpande and E.~Ma, ``{Pattern of Symmetry Breaking with Two Higgs
  Doublets},''
\href{http://dx.doi.org/10.1103/PhysRevD.18.2574}{{\em Phys. Rev.} {\bfseries
  D18} (1978) 2574}.

\bibitem{Ma:2008cu}
E.~Ma and D.~Suematsu, ``{Fermion Triplet Dark Matter and Radiative Neutrino
  Mass},'' \href{http://dx.doi.org/10.1142/S021773230903059X}{{\em Mod. Phys.
  Lett.} {\bfseries A24} (2009) 583--589},
\href{http://arxiv.org/abs/0809.0942}{{\ttfamily arXiv:0809.0942 [hep-ph]}}.

\bibitem{Araki:2011hm}
T.~Araki, C.~Q. Geng, and K.~I. Nagao, ``{Dark Matter in Inert Triplet
  Models},'' \href{http://dx.doi.org/10.1103/PhysRevD.83.075014}{{\em Phys.
  Rev.} {\bfseries D83} (2011) 075014},
\href{http://arxiv.org/abs/1102.4906}{{\ttfamily arXiv:1102.4906 [hep-ph]}}.

\bibitem{Ma:2006km}
E.~Ma, ``{Verifiable radiative seesaw mechanism of neutrino mass and dark
  matter},'' \href{http://dx.doi.org/10.1103/PhysRevD.73.077301}{{\em Phys.
  Rev.} {\bfseries D73} (2006) 077301},
\href{http://arxiv.org/abs/hep-ph/0601225}{{\ttfamily arXiv:hep-ph/0601225
  [hep-ph]}}.

\bibitem{Cohen:2011ec}
T.~Cohen, J.~Kearney, A.~Pierce, and D.~Tucker-Smith, ``{Singlet-Doublet Dark
  Matter},'' \href{http://dx.doi.org/10.1103/PhysRevD.85.075003}{{\em Phys.
  Rev.} {\bfseries D85} (2012) 075003},
\href{http://arxiv.org/abs/1109.2604}{{\ttfamily arXiv:1109.2604 [hep-ph]}}.

\bibitem{Aprile:2018dbl}
{\bfseries XENON} Collaboration, E.~Aprile {\em et~al.}, ``{Dark Matter Search
  Results from a One Ton-Year Exposure of XENON1T},''
  \href{http://dx.doi.org/10.1103/PhysRevLett.121.111302}{{\em Phys. Rev.
  Lett.} {\bfseries 121} no.~11, (2018) 111302},
\href{http://arxiv.org/abs/1805.12562}{{\ttfamily arXiv:1805.12562
  [astro-ph.CO]}}.

\bibitem{Vergados:2008jp}
J.~D. Vergados and H.~Ejiri, ``{Can Solar Neutrinos be a Serious Background in
  Direct Dark Matter Searches?},''
  \href{http://dx.doi.org/10.1016/j.nuclphysb.2008.06.004}{{\em Nucl. Phys.}
  {\bfseries B804} (2008) 144--159},
\href{http://arxiv.org/abs/0805.2583}{{\ttfamily arXiv:0805.2583 [hep-ph]}}.

\bibitem{Boehm:2018sux}
C.~Bœhm, D.~G. Cerdeño, P.~A.~N. Machado, A.~Olivares-Del~Campo, E.~Perdomo,
  and E.~Reid, ``{How high is the neutrino floor?},''
  \href{http://dx.doi.org/10.1088/1475-7516/2019/01/043}{{\em JCAP} {\bfseries
  1901} no.~01, (2019) 043},
\href{http://arxiv.org/abs/1809.06385}{{\ttfamily arXiv:1809.06385 [hep-ph]}}.

\bibitem{Camargo:2019ukv}
D.~A. Camargo, M.~D. Campos, T.~B. de~Melo, and F.~S. Queiroz, ``{A Two Higgs
  Doublet Model for Dark Matter and Neutrino Masses},''
  \href{http://dx.doi.org/10.1016/j.physletb.2019.06.020}{{\em Phys. Lett.}
  {\bfseries B795} (2019) 319--326},
\href{http://arxiv.org/abs/1901.05476}{{\ttfamily arXiv:1901.05476 [hep-ph]}}.

\bibitem{Restrepo:2015ura}
D.~Restrepo, A.~Rivera, M.~Sánchez-Peláez, O.~Zapata, and W.~Tangarife,
  ``{Radiative Neutrino Masses in the Singlet-Doublet Fermion Dark Matter Model
  with Scalar Singlets},''
  \href{http://dx.doi.org/10.1103/PhysRevD.92.013005}{{\em Phys. Rev.}
  {\bfseries D92} no.~1, (2015) 013005},
\href{http://arxiv.org/abs/1504.07892}{{\ttfamily arXiv:1504.07892 [hep-ph]}}.

\bibitem{Ahriche:2017iar}
A.~Ahriche, A.~Jueid, and S.~Nasri, ``{Radiative neutrino mass and Majorana
  dark matter within an inert Higgs doublet model},''
  \href{http://dx.doi.org/10.1103/PhysRevD.97.095012}{{\em Phys. Rev.}
  {\bfseries D97} no.~9, (2018) 095012},
\href{http://arxiv.org/abs/1710.03824}{{\ttfamily arXiv:1710.03824 [hep-ph]}}.

\bibitem{Fiaschi:2018rky}
J.~Fiaschi, M.~Klasen, and S.~May, ``{Singlet-doublet fermion and triplet
  scalar dark matter with radiative neutrino masses},''
  \href{http://dx.doi.org/10.1007/JHEP05(2019)015}{{\em JHEP} {\bfseries 05}
  (2019) 015},
\href{http://arxiv.org/abs/1812.11133}{{\ttfamily arXiv:1812.11133 [hep-ph]}}.

\bibitem{Baumholzer:2018sfb}
S.~Baumholzer, V.~Brdar, and P.~Schwaller, ``{The New $\nu$MSM ($\nu\nu$MSM):
  Radiative Neutrino Masses, keV-Scale Dark Matter and Viable Leptogenesis with
  sub-TeV New Physics},'' \href{http://dx.doi.org/10.1007/JHEP08(2018)067}{{\em
  JHEP} {\bfseries 08} (2018) 067},
\href{http://arxiv.org/abs/1806.06864}{{\ttfamily arXiv:1806.06864 [hep-ph]}}.

\bibitem{Baumholzer:2019twf}
S.~Baumholzer, V.~Brdar, P.~Schwaller, and A.~Segner, ``{Shining Light on the
  Scotogenic Model: Interplay of Colliders, Cosmology and Astrophysics},''
\href{http://arxiv.org/abs/1912.08215}{{\ttfamily arXiv:1912.08215 [hep-ph]}}.

\bibitem{Bhattacharya:2017sml}
S.~Bhattacharya, N.~Sahoo, and N.~Sahu, ``{Singlet-Doublet Fermionic Dark
  Matter, Neutrino Mass and Collider Signatures},''
  \href{http://dx.doi.org/10.1103/PhysRevD.96.035010}{{\em Phys. Rev. D}
  {\bfseries 96} no.~3, (2017) 035010},
  \href{http://arxiv.org/abs/1704.03417}{{\ttfamily arXiv:1704.03417
  [hep-ph]}}.

\bibitem{Das:2019ntw}
P.~Das, M.~K. Das, and N.~Khan, ``{Phenomenological study of neutrino mass,
  dark matter and baryogenesis within the framework of minimal extended
  seesaw},'' \href{http://dx.doi.org/10.1007/JHEP03(2020)018}{{\em JHEP}
  {\bfseries 03} (2020) 018}, \href{http://arxiv.org/abs/1911.07243}{{\ttfamily
  arXiv:1911.07243 [hep-ph]}}.

\bibitem{Kashiwase:2015pra}
S.~Kashiwase, H.~Okada, Y.~Orikasa, and T.~Toma, ``{Two Loop Neutrino Model
  with Dark Matter and Leptogenesis},''
  \href{http://dx.doi.org/10.1142/S0217751X16501219}{{\em Int. J. Mod. Phys.}
  {\bfseries A31} no.~20n21, (2016) 1650121},
\href{http://arxiv.org/abs/1505.04665}{{\ttfamily arXiv:1505.04665 [hep-ph]}}.

\bibitem{GonzalezFelipe:2003fi}
R.~Gonzalez~Felipe, F.~R. Joaquim, and B.~M. Nobre, ``{Radiatively induced
  leptogenesis in a minimal seesaw model},''
  \href{http://dx.doi.org/10.1103/PhysRevD.70.085009}{{\em Phys. Rev.}
  {\bfseries D70} (2004) 085009},
\href{http://arxiv.org/abs/hep-ph/0311029}{{\ttfamily arXiv:hep-ph/0311029
  [hep-ph]}}.

\bibitem{Fraser:2014yha}
S.~Fraser, E.~Ma, and O.~Popov, ``{Scotogenic Inverse Seesaw Model of Neutrino
  Mass},'' \href{http://dx.doi.org/10.1016/j.physletb.2014.08.069}{{\em Phys.
  Lett.} {\bfseries B737} (2014) 280--282},
\href{http://arxiv.org/abs/1408.4785}{{\ttfamily arXiv:1408.4785 [hep-ph]}}.

\bibitem{Merle:2015ica}
A.~Merle and M.~Platscher, ``{Running of radiative neutrino masses: the
  scotogenic model — revisited},''
  \href{http://dx.doi.org/10.1007/JHEP11(2015)148}{{\em JHEP} {\bfseries 11}
  (2015) 148},
\href{http://arxiv.org/abs/1507.06314}{{\ttfamily arXiv:1507.06314 [hep-ph]}}.

\bibitem{Law:2013saa}
S.~S.~C. Law and K.~L. McDonald, ``{A Class of Inert N-tuplet Models with
  Radiative Neutrino Mass and Dark Matter},''
  \href{http://dx.doi.org/10.1007/JHEP09(2013)092}{{\em JHEP} {\bfseries 09}
  (2013) 092},
\href{http://arxiv.org/abs/1305.6467}{{\ttfamily arXiv:1305.6467 [hep-ph]}}.

\bibitem{Mahanta:2019gfe}
D.~Mahanta and D.~Borah, ``{Fermion Dark Matter with $N_2$ Leptogenesis in
  Minimal Scotogenic Model},''
  \href{http://dx.doi.org/10.1088/1475-7516/2019/11/021}{{\em JCAP} {\bfseries
  1911} no.~11, (2019) 021},
\href{http://arxiv.org/abs/1906.03577}{{\ttfamily arXiv:1906.03577 [hep-ph]}}.

\bibitem{Klein:2019iws}
C.~Klein, M.~Lindner, and S.~Ohmer, ``{Minimal Radiative Neutrino Masses},''
  \href{http://dx.doi.org/10.1007/JHEP03(2019)018}{{\em JHEP} {\bfseries 03}
  (2019) 018},
\href{http://arxiv.org/abs/1901.03225}{{\ttfamily arXiv:1901.03225 [hep-ph]}}.

\bibitem{Bennett:2006fi}
{\bfseries Muon g-2} Collaboration, G.~Bennett {\em et~al.}, ``{Final Report of
  the Muon E821 Anomalous Magnetic Moment Measurement at BNL},''
  \href{http://dx.doi.org/10.1103/PhysRevD.73.072003}{{\em Phys. Rev. D}
  {\bfseries 73} (2006) 072003},
  \href{http://arxiv.org/abs/hep-ex/0602035}{{\ttfamily arXiv:hep-ex/0602035}}.

\bibitem{Parker:2018vye}
R.~H. Parker, C.~Yu, W.~Zhong, B.~Estey, and H.~M\"uller, ``{Measurement of the
  fine-structure constant as a test of the Standard Model},''
  \href{http://dx.doi.org/10.1126/science.aap7706}{{\em Science} {\bfseries
  360} (2018) 191}, \href{http://arxiv.org/abs/1812.04130}{{\ttfamily
  arXiv:1812.04130 [physics.atom-ph]}}.

\bibitem{Abe:2017jqo}
T.~Abe, R.~Sato, and K.~Yagyu, ``{Muon specific two-Higgs-doublet model},''
  \href{http://dx.doi.org/10.1007/JHEP07(2017)012}{{\em JHEP} {\bfseries 07}
  (2017) 012}, \href{http://arxiv.org/abs/1705.01469}{{\ttfamily
  arXiv:1705.01469 [hep-ph]}}.

\bibitem{Chun:2016hzs}
E.~J. Chun and J.~Kim, ``{Leptonic Precision Test of Leptophilic
  Two-Higgs-Doublet Model},''
  \href{http://dx.doi.org/10.1007/JHEP07(2016)110}{{\em JHEP} {\bfseries 07}
  (2016) 110}, \href{http://arxiv.org/abs/1605.06298}{{\ttfamily
  arXiv:1605.06298 [hep-ph]}}.

\bibitem{Baek:2001kca}
S.~Baek, N.~Deshpande, X.~He, and P.~Ko, ``{Muon anomalous g-2 and gauged
  L(muon) - L(tau) models},''
  \href{http://dx.doi.org/10.1103/PhysRevD.64.055006}{{\em Phys. Rev. D}
  {\bfseries 64} (2001) 055006},
  \href{http://arxiv.org/abs/hep-ph/0104141}{{\ttfamily arXiv:hep-ph/0104141}}.

\bibitem{Endo:2012hp}
M.~Endo, K.~Hamaguchi, and G.~Mishima, ``{Constraints on Hidden Photon Models
  from Electron g-2 and Hydrogen Spectroscopy},''
  \href{http://dx.doi.org/10.1103/PhysRevD.86.095029}{{\em Phys. Rev. D}
  {\bfseries 86} (2012) 095029},
  \href{http://arxiv.org/abs/1209.2558}{{\ttfamily arXiv:1209.2558 [hep-ph]}}.

\bibitem{Abe:2019bkf}
Y.~Abe, T.~Toma, and K.~Tsumura, ``{A $\mu$-$\tau$-philic scalar doublet under
  $Z_n$ flavor symmetry},''
  \href{http://dx.doi.org/10.1007/JHEP06(2019)142}{{\em JHEP} {\bfseries 06}
  (2019) 142}, \href{http://arxiv.org/abs/1904.10908}{{\ttfamily
  arXiv:1904.10908 [hep-ph]}}.

\bibitem{Bhattacharya:2018cgx}
S.~Bhattacharya, P.~Ghosh, and N.~Sahu, ``{Multipartite Dark Matter with
  Scalars, Fermions and signatures at LHC},''
  \href{http://dx.doi.org/10.1007/JHEP02(2019)059}{{\em JHEP} {\bfseries 02}
  (2019) 059}, \href{http://arxiv.org/abs/1809.07474}{{\ttfamily
  arXiv:1809.07474 [hep-ph]}}.

\bibitem{Cynolter:2008ea}
G.~Cynolter and E.~Lendvai, ``{Electroweak Precision Constraints on Vector-like
  Fermions},'' \href{http://dx.doi.org/10.1140/epjc/s10052-008-0771-7}{{\em
  Eur. Phys. J.} {\bfseries C58} (2008) 463--469},
\href{http://arxiv.org/abs/0804.4080}{{\ttfamily arXiv:0804.4080 [hep-ph]}}.

\bibitem{Gu:2018kmv}
P.-H. Gu and H.-J. He, ``{TeV Scale Neutrino Mass Generation, Minimal Inelastic
  Dark Matter, and High Scale Leptogenesis},''
  \href{http://dx.doi.org/10.1103/PhysRevD.99.015025}{{\em Phys. Rev. D}
  {\bfseries 99} no.~1, (2019) 015025},
  \href{http://arxiv.org/abs/1808.09377}{{\ttfamily arXiv:1808.09377
  [hep-ph]}}.

\bibitem{ano1}
P.~Pal.

\bibitem{ano2}
F.~Pisano and A.~T. Tran.

\bibitem{Pal:1690642}
P.~B. Pal, \href{http://dx.doi.org/1482216981}{{\em {An introductory course of
  particle physics}}}.
\newblock Boca Raton, FL, Jul, 2014.
\newblock \url{https://cds.cern.ch/record/1690642}.

\bibitem{Kannike:2016fmd}
K.~Kannike, ``{Vacuum Stability of a General Scalar Potential of a Few
  Fields},'' \href{http://dx.doi.org/10.1140/epjc/s10052-016-4160-3}{{\em Eur.
  Phys. J. C} {\bfseries 76} no.~6, (2016) 324},
  \href{http://arxiv.org/abs/1603.02680}{{\ttfamily arXiv:1603.02680
  [hep-ph]}}. [Erratum: Eur.Phys.J.C 78, 355 (2018)].

\bibitem{Garg:2017iva}
I.~Garg, S.~Goswami, K.~N. Vishnudath, and N.~Khan, ``{Electroweak vacuum
  stability in presence of singlet scalar dark matter in TeV scale seesaw
  models},'' \href{http://dx.doi.org/10.1103/PhysRevD.96.055020}{{\em Phys.
  Rev.} {\bfseries D96} no.~5, (2017) 055020},
\href{http://arxiv.org/abs/1706.08851}{{\ttfamily arXiv:1706.08851 [hep-ph]}}.

\bibitem{Khan:2012zw}
S.~Khan, S.~Goswami, and S.~Roy, ``{Vacuum Stability constraints on the minimal
  singlet TeV Seesaw Model},''
  \href{http://dx.doi.org/10.1103/PhysRevD.89.073021}{{\em Phys. Rev.}
  {\bfseries D89} no.~7, (2014) 073021},
\href{http://arxiv.org/abs/1212.3694}{{\ttfamily arXiv:1212.3694 [hep-ph]}}.

\bibitem{Lee:1977eg}
B.~W. Lee, C.~Quigg, and H.~B. Thacker, ``{Weak Interactions at Very
  High-Energies: The Role of the Higgs Boson Mass},''
\href{http://dx.doi.org/10.1103/PhysRevD.16.1519}{{\em Phys. Rev.} {\bfseries
  D16} (1977) 1519}.

\bibitem{Cynolter:2004cq}
G.~Cynolter, E.~Lendvai, and G.~Pocsik, ``{Note on unitarity constraints in a
  model for a singlet scalar dark matter candidate},'' {\em Acta Phys. Polon.}
  {\bfseries B36} (2005) 827--832,
\href{http://arxiv.org/abs/hep-ph/0410102}{{\ttfamily arXiv:hep-ph/0410102
  [hep-ph]}}.

\bibitem{Djouadi:2005gj}
A.~Djouadi, ``{The Anatomy of electro-weak symmetry breaking. II. The Higgs
  bosons in the minimal supersymmetric model},''
  \href{http://dx.doi.org/10.1016/j.physrep.2007.10.005}{{\em Phys. Rept.}
  {\bfseries 459} (2008) 1--241},
\href{http://arxiv.org/abs/hep-ph/0503173}{{\ttfamily arXiv:hep-ph/0503173
  [hep-ph]}}.

\bibitem{Baak:2014ora}
{\bfseries Gfitter Group} Collaboration, M.~Baak, J.~Cúth, J.~Haller,
  A.~Hoecker, R.~Kogler, K.~Mönig, M.~Schott, and J.~Stelzer, ``{The global
  electroweak fit at NNLO and prospects for the LHC and ILC},''
  \href{http://dx.doi.org/10.1140/epjc/s10052-014-3046-5}{{\em Eur. Phys. J.}
  {\bfseries C74} (2014) 3046},
\href{http://arxiv.org/abs/1407.3792}{{\ttfamily arXiv:1407.3792 [hep-ph]}}.

\bibitem{Peskin:1991sw}
M.~E. Peskin and T.~Takeuchi, ``{Estimation of oblique electroweak
  corrections},''
\href{http://dx.doi.org/10.1103/PhysRevD.46.381}{{\em Phys. Rev.} {\bfseries
  D46} (1992) 381--409}.

\bibitem{Aghanim:2018eyx}
{\bfseries Planck} Collaboration, N.~Aghanim {\em et~al.}, ``{Planck 2018
  results. VI. Cosmological parameters},''
  \href{http://arxiv.org/abs/1807.06209}{{\ttfamily arXiv:1807.06209
  [astro-ph.CO]}}.

\bibitem{fermilat1}
{\bfseries Fermi LAT collaboration} Collaboration, M.~A. et~al., ``{Measurement
  of separate cosmic-ray electron and positron spectra with the Fermi Large
  Area Telescope},''
  \href{http://dx.doi.org/10.1103/PhysRevLett.108.011103}{{\em Phys. Rev.
  Lett.} {\bfseries 108} (2012) 011103},
\href{http://arxiv.org/abs/1109.0521}{{\ttfamily arXiv:1109.0521 [hep-ex]}}.

\bibitem{Athron:2017kgt}
{\bfseries GAMBIT} Collaboration, P.~Athron {\em et~al.}, ``{Status of the
  scalar singlet dark matter model},''
  \href{http://dx.doi.org/10.1140/epjc/s10052-017-5113-1}{{\em Eur. Phys. J.}
  {\bfseries C77} no.~8, (2017) 568},
\href{http://arxiv.org/abs/1705.07931}{{\ttfamily arXiv:1705.07931 [hep-ph]}}.

\bibitem{indirect2}
J.~S. Michael~Duerr, Pavel Fileviez~Perez, ``{Scalar Dark Matter: Direct vs.
  Indirect Detection},'' \href{http://dx.doi.org/10.1007/JHEP06(2016)152}{{\em
  JHEP} {\bfseries 06} (2016) 152},
\href{http://arxiv.org/abs/1509.04282}{{\ttfamily arXiv:1509.04282 [hep-ph]}}.

\bibitem{Alloul:2013bka}
A.~Alloul, N.~D. Christensen, C.~Degrande, C.~Duhr, and B.~Fuks, ``{FeynRules
  2.0 - A complete toolbox for tree-level phenomenology},''
  \href{http://dx.doi.org/10.1016/j.cpc.2014.04.012}{{\em Comput. Phys.
  Commun.} {\bfseries 185} (2014) 2250--2300},
\href{http://arxiv.org/abs/1310.1921}{{\ttfamily arXiv:1310.1921 [hep-ph]}}.

\bibitem{Belanger:2018mqt}
G.~Bélanger, F.~Boudjema, A.~Goudelis, A.~Pukhov, and B.~Zaldivar,
  ``{micrOMEGAs5.0 : Freeze-in},''
  \href{http://dx.doi.org/10.1016/j.cpc.2018.04.027}{{\em Comput. Phys.
  Commun.} {\bfseries 231} (2018) 173--186},
\href{http://arxiv.org/abs/1801.03509}{{\ttfamily arXiv:1801.03509 [hep-ph]}}.

\bibitem{Baldini:2018nnn}
{\bfseries MEG II} Collaboration, A.~M. Baldini {\em et~al.}, ``{The design of
  the MEG II experiment},''
  \href{http://dx.doi.org/10.1140/epjc/s10052-018-5845-6}{{\em Eur. Phys. J.}
  {\bfseries C78} no.~5, (2018) 380},
\href{http://arxiv.org/abs/1801.04688}{{\ttfamily arXiv:1801.04688
  [physics.ins-det]}}.

\bibitem{Harnik:2012pb}
R.~Harnik, J.~Kopp, and J.~Zupan, ``{Flavor Violating Higgs Decays},''
  \href{http://dx.doi.org/10.1007/JHEP03(2013)026}{{\em JHEP} {\bfseries 03}
  (2013) 026}, \href{http://arxiv.org/abs/1209.1397}{{\ttfamily arXiv:1209.1397
  [hep-ph]}}.

\bibitem{FileviezPerez:2009ud}
P.~Fileviez~Perez and M.~B. Wise, ``{On the Origin of Neutrino Masses},''
  \href{http://dx.doi.org/10.1103/PhysRevD.80.053006}{{\em Phys. Rev.}
  {\bfseries D80} (2009) 053006},
\href{http://arxiv.org/abs/0906.2950}{{\ttfamily arXiv:0906.2950 [hep-ph]}}.

\bibitem{tHooft:1980xss}
G.~'t~Hooft, C.~Itzykson, A.~Jaffe, H.~Lehmann, P.~K. Mitter, I.~M. Singer, and
  R.~Stora, ``{Recent Developments in Gauge Theories. Proceedings, Nato
  Advanced Study Institute, Cargese, France, August 26 - September 8, 1979},''
\href{http://dx.doi.org/10.1007/978-1-4684-7571-5}{{\em NATO Sci. Ser. B}
  {\bfseries 59} (1980) pp.1--438}.

\bibitem{McDonald:1993ex}
J.~McDonald, ``{Gauge singlet scalars as cold dark matter},''
  \href{http://dx.doi.org/10.1103/PhysRevD.50.3637}{{\em Phys. Rev.} {\bfseries
  D50} (1994) 3637--3649},
\href{http://arxiv.org/abs/hep-ph/0702143}{{\ttfamily arXiv:hep-ph/0702143
  [HEP-PH]}}.

\bibitem{Griest:1990kh}
K.~Griest and D.~Seckel, ``{Three exceptions in the calculation of relic
  abundances},''
\href{http://dx.doi.org/10.1103/PhysRevD.43.3191}{{\em Phys. Rev.} {\bfseries
  D43} (1991) 3191--3203}.

\bibitem{Giusarma:2016phn}
E.~Giusarma, M.~Gerbino, O.~Mena, S.~Vagnozzi, S.~Ho, and K.~Freese,
  ``{Improvement of cosmological neutrino mass bounds},''
  \href{http://dx.doi.org/10.1103/PhysRevD.94.083522}{{\em Phys. Rev.}
  {\bfseries D94} no.~8, (2016) 083522},
\href{http://arxiv.org/abs/1605.04320}{{\ttfamily arXiv:1605.04320
  [astro-ph.CO]}}.

\bibitem{Vagnozzi:2017ovm}
S.~Vagnozzi, E.~Giusarma, O.~Mena, K.~Freese, M.~Gerbino, S.~Ho, and
  M.~Lattanzi, ``{Unveiling $\nu$ secrets with cosmological data: neutrino
  masses and mass hierarchy},''
  \href{http://dx.doi.org/10.1103/PhysRevD.96.123503}{{\em Phys. Rev.}
  {\bfseries D96} no.~12, (2017) 123503},
\href{http://arxiv.org/abs/1701.08172}{{\ttfamily arXiv:1701.08172
  [astro-ph.CO]}}.

\bibitem{Casas:2001sr}
J.~Casas and A.~Ibarra, ``{Oscillating neutrinos and $\mu \to e, \gamma$},''
  \href{http://dx.doi.org/10.1016/S0550-3213(01)00475-8}{{\em Nucl. Phys. B}
  {\bfseries 618} (2001) 171--204},
  \href{http://arxiv.org/abs/hep-ph/0103065}{{\ttfamily arXiv:hep-ph/0103065}}.

\bibitem{Toma:2013zsa}
T.~Toma and A.~Vicente, ``{Lepton Flavor Violation in the Scotogenic Model},''
  \href{http://dx.doi.org/10.1007/JHEP01(2014)160}{{\em JHEP} {\bfseries 01}
  (2014) 160}, \href{http://arxiv.org/abs/1312.2840}{{\ttfamily arXiv:1312.2840
  [hep-ph]}}.

\bibitem{Esteban:2020cvm}
I.~Esteban, M.~Gonzalez-Garcia, M.~Maltoni, T.~Schwetz, and A.~Zhou, ``{The
  fate of hints: updated global analysis of three-flavor neutrino
  oscillations},'' \href{http://dx.doi.org/10.1007/JHEP09(2020)178}{{\em JHEP}
  {\bfseries 09} (2020) 178}, \href{http://arxiv.org/abs/2007.14792}{{\ttfamily
  arXiv:2007.14792 [hep-ph]}}.

\bibitem{Alwall:2014hca}
J.~Alwall, R.~Frederix, S.~Frixione, V.~Hirschi, F.~Maltoni, O.~Mattelaer,
  H.~S. Shao, T.~Stelzer, P.~Torrielli, and M.~Zaro, ``{The automated
  computation of tree-level and next-to-leading order differential cross
  sections, and their matching to parton shower simulations},''
  \href{http://dx.doi.org/10.1007/JHEP07(2014)079}{{\em JHEP} {\bfseries 07}
  (2014) 079}, \href{http://arxiv.org/abs/1405.0301}{{\ttfamily arXiv:1405.0301
  [hep-ph]}}.

\bibitem{Staub:2012pb}
F.~Staub, ``{SARAH 3.2: Dirac Gauginos, UFO output, and more},''
  \href{http://dx.doi.org/10.1016/j.cpc.2013.02.019}{{\em Comput. Phys.
  Commun.} {\bfseries 184} (2013) 1792--1809},
  \href{http://arxiv.org/abs/1207.0906}{{\ttfamily arXiv:1207.0906 [hep-ph]}}.

\bibitem{Staub:2015kfa}
F.~Staub, ``{Exploring new models in all detail with SARAH},''
  \href{http://dx.doi.org/10.1155/2015/840780}{{\em Adv. High Energy Phys.}
  {\bfseries 2015} (2015) 840780},
  \href{http://arxiv.org/abs/1503.04200}{{\ttfamily arXiv:1503.04200
  [hep-ph]}}.

\bibitem{Porod:2011nf}
W.~Porod and F.~Staub, ``{SPheno 3.1: Extensions including flavour, CP-phases
  and models beyond the MSSM},''
  \href{http://dx.doi.org/10.1016/j.cpc.2012.05.021}{{\em Comput. Phys.
  Commun.} {\bfseries 183} (2012) 2458--2469},
  \href{http://arxiv.org/abs/1104.1573}{{\ttfamily arXiv:1104.1573 [hep-ph]}}.

\bibitem{Sjostrand:2014zea}
T.~Sj\"ostrand, S.~Ask, J.~R. Christiansen, R.~Corke, N.~Desai, P.~Ilten,
  S.~Mrenna, S.~Prestel, C.~O. Rasmussen, and P.~Z. Skands, ``{An introduction
  to PYTHIA 8.2},'' \href{http://dx.doi.org/10.1016/j.cpc.2015.01.024}{{\em
  Comput. Phys. Commun.} {\bfseries 191} (2015) 159--177},
  \href{http://arxiv.org/abs/1410.3012}{{\ttfamily arXiv:1410.3012 [hep-ph]}}.

\bibitem{deFavereau:2013fsa}
{\bfseries DELPHES 3} Collaboration, J.~de~Favereau, C.~Delaere, P.~Demin,
  A.~Giammanco, V.~Lema\^\i{}tre, A.~Mertens, and M.~Selvaggi, ``{DELPHES 3, A
  modular framework for fast simulation of a generic collider experiment},''
  \href{http://dx.doi.org/10.1007/JHEP02(2014)057}{{\em JHEP} {\bfseries 02}
  (2014) 057}, \href{http://arxiv.org/abs/1307.6346}{{\ttfamily arXiv:1307.6346
  [hep-ex]}}.

\bibitem{Barman:2020azo}
R.~K. Barman, B.~Bhattacherjee, I.~Chakraborty, A.~Choudhury, and N.~Khan,
  ``{Electroweakino searches at the HL-LHC in the baryon number violating
  MSSM},'' \href{http://arxiv.org/abs/2003.10920}{{\ttfamily arXiv:2003.10920
  [hep-ph]}}.

\bibitem{Lester:1999tx}
C.~Lester and D.~Summers, ``{Measuring masses of semiinvisibly decaying
  particles pair produced at hadron colliders},''
  \href{http://dx.doi.org/10.1016/S0370-2693(99)00945-4}{{\em Phys. Lett. B}
  {\bfseries 463} (1999) 99--103},
  \href{http://arxiv.org/abs/hep-ph/9906349}{{\ttfamily arXiv:hep-ph/9906349}}.

\bibitem{Cho:2007qv}
W.~S. Cho, K.~Choi, Y.~G. Kim, and C.~B. Park, ``{Gluino Stransverse Mass},''
  \href{http://dx.doi.org/10.1103/PhysRevLett.100.171801}{{\em Phys. Rev.
  Lett.} {\bfseries 100} (2008) 171801},
  \href{http://arxiv.org/abs/0709.0288}{{\ttfamily arXiv:0709.0288 [hep-ph]}}.

\bibitem{Aad:2019vnb}
{\bfseries ATLAS} Collaboration, G.~Aad {\em et~al.}, ``{Search for electroweak
  production of charginos and sleptons decaying into final states with two
  leptons and missing transverse momentum in $\sqrt{s}=13$ TeV $pp$ collisions
  using the ATLAS detector},''
  \href{http://dx.doi.org/10.1140/epjc/s10052-019-7594-6}{{\em Eur. Phys. J. C}
  {\bfseries 80} no.~2, (2020) 123},
  \href{http://arxiv.org/abs/1908.08215}{{\ttfamily arXiv:1908.08215
  [hep-ex]}}.

\bibitem{Griest:1989wd}
K.~Griest and M.~Kamionkowski, ``{Unitarity Limits on the Mass and Radius of
  Dark Matter Particles},''
\href{http://dx.doi.org/10.1103/PhysRevLett.64.615}{{\em Phys. Rev. Lett.}
  {\bfseries 64} (1990) 615}.

\bibitem{Smirnov:2019ngs}
J.~Smirnov and J.~F. Beacom, ``{TeV-Scale Thermal WIMPs: Unitarity and its
  Consequences},'' \href{http://dx.doi.org/10.1103/PhysRevD.100.043029}{{\em
  Phys. Rev.} {\bfseries D100} no.~4, (2019) 043029},
\href{http://arxiv.org/abs/1904.11503}{{\ttfamily arXiv:1904.11503 [hep-ph]}}.

\end{thebibliography}\endgroup
\end{document}